\documentclass[aps,prb,superscriptaddress,amsmath,amssymb,showpacs,twocolumn]{revtex4-1}

\usepackage{graphicx}
\usepackage{hyperref}
\usepackage{epsfig}
\usepackage{color}
\usepackage{psfrag}
\usepackage{float}
\usepackage{bbold}
\usepackage{tikz}

\newcommand{\ket}[1] {\left|#1\right\rangle}
\newcommand{\bra}[1] {\langle #1 |}

\newcommand{\rem}[1]{}

\newcommand{\refe}[1]{~(\ref{#1})}
\newcommand{\Eqref}[1]{Eq.~(\ref{#1})}

\begin{document}

\title{Bistability and Displacement Fluctuations in a Quantum Nano-mechanical Oscillator}

\author{R. Avriller} 
\affiliation{Univ. Bordeaux, CNRS, LOMA, UMR 5798, F-33405 Talence, France}
\author{B. Murr} 
\affiliation{Univ. Bordeaux, CNRS, LOMA, UMR 5798, F-33405 Talence, France}
\author{F. Pistolesi} 
\affiliation{Univ. Bordeaux, CNRS, LOMA, UMR 5798, F-33405 Talence, France}
\date{\today}

\begin{abstract}
Remarkable features have been predicted for the mechanical fluctuations at the bistability transition of a classical oscillator coupled capacitively to a quantum dot 
[Phys. Rev. Lett. \textbf{115}, 206802 (2015)].
These results have been obtained in the regime $\hbar\omega_0 \ll k_B T \ll \hbar\Gamma$, where $\omega_0$, $T$, and $\Gamma$ are the mechanical resonating frequency, 
the temperature, and the tunneling rate, respectively.
A similar behavior could be expected in the quantum regime of $\hbar\Gamma \ll k_B T \ll \hbar\omega_0$.
We thus calculate the energy and displacement fluctuation spectra and study their behavior as a function of the electro-mechanical coupling constant
when the system enters the Frank-Condon regime. 
We find that, in analogy with the classical case, the energy fluctuation spectrum and the displacement spectrum
widths show a maximum for values of the coupling constant at which a mechanical bistability 
establishes.
\end{abstract}

\maketitle

%
\section{Introduction}
Nano-electromechanical systems (NEMS) have proved to be devices of great interest, both from
fundamental and applicative point of views \cite{IMBODEN201489}.
A paradigmatic example of such devices is represented by suspended carbon-nanotube mechanical 
resonators \cite{sazonova_tunable_2004,lassagne_coupling_2009,steele_strong_2009,SMLL:SMLL200901984}.
Due to their low mass ($10^{-18}$ g) and high Young modulus (1 TPa), carbon nanotube
mechanical oscillators are ideal candidates for developing a new generation of ultra-sensitive force and mass sensors.
A lot of efforts were thus devoted in the past decades in order to propose efficient
schemes to actuate and detect the mechanical motion of such devices.
The mixing technique is one of those approaches \cite{sazonova_tunable_2004,SMLL:SMLL200901984}.
Initially proposed in Ref.~\onlinecite{sazonova_tunable_2004}, it enables to excite mechanically a
nanotube quantum dot by applying suitable time-dependent
gate and bias voltages.
The resulting mechanical oscillation of the nanotube in the frequency range
$\omega_0/2\pi\approx 100 \mbox{ MHz}-10 \mbox{ GHz}$ \cite{laird_high_2011,High_frequency_nanotube_mechanical_resonators} is then
transduced toward a measurable lower-frequency electronic mixing-current.
The later contains informations about both quadratures of the nanotube displacement and thus about its mechanical susceptibility.
This technique was used to measure tiny variations of the resonance frequency in real time, upon adsorption of molecules on the surface of the nanotube \cite{chaste_nanomechanical_2012}.
This enabled to perform mass-sensing experiments with a record sensitivity reported at the yoctogram resolution (proton mass) \cite{chaste_nanomechanical_2012}
and to detect the back-action of single-electron tunneling events as a measurable softening of the mechanical resonance frequency \cite{lassagne_coupling_2009,steele_strong_2009,ganzhorn_dynamics_2012,PhysRevB.86.115454}.
The optimum sensitivity achievable with the mixing-technique was investigated theoretically in 
Ref.~\onlinecite{PhysRevB.95.035410}
and was shown to arise from a compromise between maximizing the mixing signal to overcome electronic shot-noise
and minimizing the added noise corresponding to electronic back-action.
The higher the electromechanical coupling, the higher the achieved sensitivity, 
thus justifying the goal of reaching the strong-coupling regime
between tunneling electrons and one mechanical degree of freedom of the nanotube.
Recent progresses in fabrication techniques were reported that go along that direction \cite{waissman2013realization,benyamini_real-space_2014}, by
designing local quantum dots on the surface of the nanotube, with full control of their electrical and mechanical properties.
This enabled to probe regimes where the height of the tunneling barriers $\Gamma$ is either smaller or larger than $\omega_0$
as well as to image spatially the excited mechanical mode by changing the location of the quantum dot along the nanotube direction \cite{benyamini_real-space_2014}.
In those experiments, the electromechanical coupling strength is given by the polaronic energy scale
$\epsilon_P=F_0^2/k$, with $F_0$ the excess of force applied on the oscillator upon tunneling of a single electron,
and $k$ the nanotube spring constant.
Typical electromechanical coupling strengths obtained in the experiments of Ref.\cite{benyamini_real-space_2014} are
estimated from the softening of the resonance frequency to be of order $\epsilon_P \approx 0.3 \mbox{ K}$ at temperature $T = 16\mbox{ K}$
\cite{PhysRevLett.115.206802}.
Less invasive and low-noise techniques were recently proposed, the principle of which is to extract
the oscillator displacement fluctuation spectrum $S_{xx}(\omega)$ from a measurement
of the current-fluctuations across the nanotube \cite{moser2013ultrasensitive}.
Large mechanical quality factors $Q$ up to $5$ million were reported with this approach \cite{moser_nanotube_2014} as well as force sensing experiments
with a resolution up to $\approx 12 \mbox{ zN.Hz}^{-1/2}$\cite{moser2013ultrasensitive}.
Recently, some of the authors investigated theoretically measurable mechanical properties of
a classical and slow suspended carbon-nanotube \cite{PhysRevLett.115.206802,PhysRevB.94.125417},
for which $\omega_0\ll V, T \ll \Gamma$ (in the paper we use the notation that the Planck constant $\hbar$, the Boltzmann constant $k_B$ and the elementary electron charge $e$ are all set to 1).
They showed that entering the strong electromechanical coupling regime has a dramatic impact
on the oscillator displacement spectrum $S_{xx}(\omega)$.
Upon increasing $\epsilon_P/\Gamma$, the maximum frequency of the spectrum $\omega_{max}$ is softened toward lower frequencies while
the full width half maximum (FWHM) $\Delta \omega$ of the spectral line increases
up to a maximum value reached for a critical coupling strength $\epsilon_P = \pi \Gamma$.
At this critical point, the lineshape of the spectrum is dominated by a strong frequency-noise induced by
the dominating quartic non-linearities of the mechanical oscillator \cite{PhysRevLett.115.206802}.
It was predicted universal scaling behavior with bias-voltage of both $\omega_{max}$ and $\Delta\omega\approx\omega_0 \left( V/\Gamma \right)^{1/4}$
as well as a universal quality factor $Q \approx 1.7$ \cite{PhysRevLett.115.206802}.
Increasing further the electromechanical coupling $\epsilon_P > \pi \Gamma$,
the mechanical oscillator becomes effectively bistable and the electronic current across the nanotube is progressively blocked.
This phenomenon is analogous to the current-blockade transition that was predicted for a classical oscillator coupled
to incoherent tunneling electrons ($\Gamma \ll T$) when $\epsilon_P > V$ \cite{galperin_hysteresis_2005,mozyrsky_intermittent_2006,pistolesi_current_2007,pistolesi_self-consistent_2008}.
Interestingly, the critical point at which current-blockade occurs coincides with the point at which the dephasing-rate due to frequency-noise
is maximum \cite{PhysRevLett.115.206802} and the mixing technique has a maximum sensitivity \cite{0953-8984-29-46-465304}.
The full stability phase-diagram for the mechanical oscillator and the corresponding lineshapes of the position fluctuation spectra were derived as a function of bias, gate voltage and temperature in Ref.~\onlinecite{PhysRevB.94.125417}.
This effect can be observed in principle in existing samples \cite{benyamini_real-space_2014}, 
provided they are measured at temperature of the order of 20 mK.
A similar phenomenon, known as Franck-Condon blockade, has been 
predicted \cite{koch_franck-condon_2005,koch_theory_2006,braig_vibrational_2003,mitra_phonon_2004}  
and observed \cite{leturcq_franckcondon_2009,burzuri2014franck} for molecular systems 
in the opposite regime of large resonating frequency $\Gamma \ll T \ll \omega_0$,
for which the oscillator is close to its quantum ground state.
The consequences in electronic transport of the Franck-Condon blockade has been 
investigated in details, but much less is known about the dynamical properties of the mechanical oscillator
in this regime \cite{PhysRevB.83.125419,0953-8984-23-10-105301,PhysRevB.83.245311,PhysRevB.80.155437}.
The aim of the present paper is to investigate if there is a quantum counterpart of the striking behavior of the 
displacement fluctuation spectrum predicted in the classical regime: 
namely the existence and measurable manifestations of a mechanical bistability and the 
coupling constant dependence of the width $\Delta\omega$ of the displacement fluctuation spectrum.
Concerning the bistability it is well known that for strong coupling the current is blocked.
This means that electrons cannot tunnel anymore keeping the electronic dot in either the empty or full state.
Can one regard this system as a bistable one in a similar manner of the classical system?
What is the relation between the quantum displacement fluctuation spectrum and the 
appearance of the bistability?
One can anticipate that at weak coupling, $\Delta \omega$ exhibits a quadratic dependence on the coupling 
constant, coming from simple perturbative arguments, but the strong coupling limit demands more insight since
the width may have different origin: energy dissipation, classical phase fluctuations, and quantum decoherence. 
In order to answer these questions we calculate in the quantum fast oscillator regime the 
(non-symmetric) displacement spectrum, the energy fluctuation spectrum, and the 
Wigner distribution for the oscillator.
We find that the width of the energy fluctuation spectrum shows a clear maximum for the same value of the 
coupling constant for which the probability distribution develops a double peak.
This can be interpreted as the onset of the bistability.
The energy scale for this transition turns out to be $\epsilon_P = 2\omega_o$. 
The same energy scale controls the washing out of the bistability as a function of the temperature 
$\epsilon_P\approx  T$, or the voltage bias $\epsilon_P\approx  V$.
We present a detailed analytical analysis, indicating that despite 
the similarity with the classical case, the origin of the maximum of the dissipation has a different origin in the quantum case.
The behavior predicted could be observed by detecting finite frequency current noise through suspended carbon
nanotubes where electronic transport is coupled either to 
GHz flexural modes \cite{laird_high_2011,High_frequency_nanotube_mechanical_resonators} or to 
THz nanotube breathing modes \cite{leturcq_franckcondon_2009}.

The organization of the paper is the following.
In Sec.~\ref{Mechanical_system}, we introduce the microscopic Hamiltonian describing a 
mechanical oscillator coupled to a single-level quantum dot.
In Sec.~\ref{Master_equation}, we derive the generalized master-equation with the Born-Markov approximation,
that enables to compute the dynamical properties of the mechanical oscillator.
The energy and position fluctuation spectra are computed respectively in Sec.~\ref{Energy_fluctuation_spectrum}
and Sec.~\ref{Displacement_fluctuation_spectrum}.
The dissipation and decoherence mechanisms are analyzed in relation to the crossover toward bistability of the mechanical oscillator.
Finally, the bias-voltage dependence of both energy and displacement spectra are shown in
Sec.~\ref{Voltage_dependence}.
%

%
\section{The mechanical system}
\label{Mechanical_system}
%
%
\begin{figure}[tbh]
  \includegraphics[width=.7\linewidth]{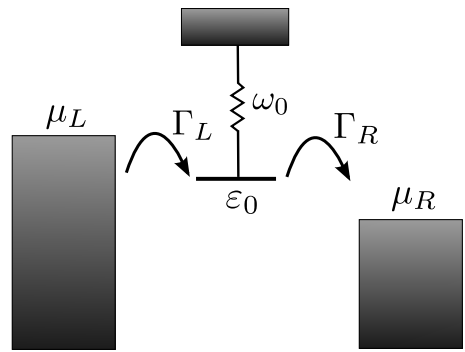}
  \caption{
    Representation of a nano-mechanical oscillator with resonance frequency $\omega_0$.
    The oscillator is coupled to a quantum dot, described by a single electronic level of
    energy $\varepsilon_0$.
    Charge is transferred from the left (right) lead to the dot with a tunneling rate $\Gamma_L$ ($\Gamma_R$).
    An externally applied bias voltage $V$ leads to a difference between the chemical potentials
    of the electronic reservoirs $\mu_L - \mu_R = V$.
  }
\label{fig:Setup}
\end{figure}
%
%
%
We consider a nano-mechanical oscillator coupled capacitively to a single-level quantum dot
[see Fig.~\ref{fig:Setup}].
Assuming spinless electrons, the microscopic Hamiltonian of the full electromechanical system is given by:
\begin{eqnarray}
  H &=& H_0 + \sum_{\alpha=L,R}H_\alpha + H_T
  \label{Hamiltonian1}\\
  H_0 &=& \left\lbrack \epsilon_0 + g \omega_0 \left( a + a^\dagger \right) \right\rbrack d^{\dagger}d 
  + \omega_0 a^{\dagger}a
  \label{Hamiltonian2}\\
  H_{\alpha} &=& \sum_{k} \left( \varepsilon_{\alpha k} - \mu_{\alpha} \right) c_{\alpha k}^{\dagger}c_{\alpha k}
  \label{Hamiltonian3}\\
  H_T &=& \sum_{\alpha=L,R}\sum_k \lbrace t_{\alpha k} c_{\alpha k}^{\dagger}d + t_{\alpha k}^* d^{\dagger}c_{\alpha k}\rbrace
  \label{Hamiltonian4}\, ,
\end{eqnarray}
where $d^\dagger$ and $a^\dagger$ are respectively the creation operator for an electron on the dot and a vibron on the mechanical oscillator.
The first term $H_0$ describes the mechanical oscillator of bare resonance frequency $\omega_0$ and
the single-level quantum dot of energy $\varepsilon_0$.
The charge operator on the dot $n_d=d^{\dagger}d$ couples linearly to the oscillator displacement operator
\begin{equation}
	x = x_0 \left( a + a^\dagger \right)
\end{equation}
with $x_0=\sqrt{1/2m\omega_0}$ its zero-point motion.
The electromechanical coupling strength in units of the vibron energy is written $g \omega_0$ with the 
excess force acting on the oscillator when one electron is added $F_0=g \omega_0/x_0$.
The second term $H_{\alpha}$ is the Hamiltonian of the $\alpha=L$ (left) and $R$ (right) free electronic reservoirs, both
characterized by an electronic band-structure $\varepsilon_{\alpha k}$ and a chemical potential $\mu_{\alpha}$.
A voltage bias $V$ is externally applied, that we will suppose to be equally shared between left and right metallic reservoirs, 
namely $\mu_L = V/2$ and $\mu_R = -V/2$.
Finally, the last term $H_T$ is the tunneling Hamiltonian.
It describes charge transfer from the electronic reservoir $\alpha=L,R$ to the quantum dot,
with a corresponding tunneling rate $\Gamma_{\alpha}=2\pi | t_{\alpha}|^2 \rho_{\alpha}$.
The former is proportional to the hopping term $t_{\alpha k}\equiv t_{\alpha}$ supposed
to be real and independent of the wave-vector $k$ and
to the electronic density of states $\rho_{\alpha}$ evaluated at the Fermi energy (wide-band approximation).
Note that the relevant energy scale of the problem is the polaronic energy defined above 
as $\epsilon_P=F_0^2/k= 2 g^2 \omega_0$.
We will see that when $\epsilon_P$ crosses the other relevant energy scales, as the temperature 
$T$, the bias voltage $V$, or the zero point motion energy $\omega_0$, the strong coupling effects 
appear to be relevant.
When only $\omega_0$ matters, one can either use $g$ or $\epsilon_P/2\omega_0=g^2$ as 
dimensionless coupling. 
We will use both in the following, since certain expressions and dependences are more transparent in terms of 
$g^2$.

We begin by performing the Lang-Firsov unitary transformation \cite{lang1963kinetic} $U = e^{gn_d\left( a - a^\dagger \right)}$ to the Hamiltonian of Eq.\refe{Hamiltonian1}.
The transformed Hamiltonian $\tilde{H}=UHU^\dagger$ is obtained as:
\begin{eqnarray}
  \tilde{H}_{0} &=& \tilde{\epsilon}_0 d^{\dagger}d + \omega_0 a^{\dagger}a 
  \label{TildeHamiltonian1} \\
  \tilde{H}_T &=& \sum_{\alpha=L,R}t_{\alpha}\sum_k \lbrace c_{\alpha k}^{\dagger} D + D^{\dagger}c_{\alpha k}\rbrace
  \label{TildeHamiltonian2} \, .
\end{eqnarray}
The meaning of Eq.\refe{TildeHamiltonian1} is the following:
upon tunneling of a single electron, the quantum dot is excited into a charged electronic state.
The corresponding excess energy can be partially released by relaxation of the mechanical oscillator
into a new equilibrium position $\tilde{X}_{eq} = - 2g x_{0}$.
The energy of the single-level quantum dot $\tilde{\epsilon}_0 = \epsilon_0 - \epsilon_P/2$
is consequently reduced by the polaronic shift.
Any explicit term involving the electromechanical coupling has thus disappeared
from the expression of $\tilde{H}_{0}$, at the price of modifying the tunneling 
Hamiltonian Eq.\refe{TildeHamiltonian2}.
The hopping terms $t_\alpha$ belonging to $\tilde{H}_T$ are renormalized by the polaron cloud operator
$Q=e^{g\left( a - a^\dagger \right)}$ and incorporated into a redefinition of the dot annihilation 
operator $D \equiv d\, Q$.
The displacement operator is modified also by the same transformation and can be written:
\begin{equation}
	x\rightarrow U x U^\dag = X-2 g n_d x_0
	\label{xdef}
\end{equation}
where $X=x_0 (a+a^\dag)$ and the dynamics of the operators $a$ and $n_d$ is now ruled by $\tilde H$.

In the following, we consider the regime of electron incoherent transport and quantum oscillator.
This regime is achieved when the reservoir temperature $T$ is larger than the total tunneling rate
$\Gamma = \Gamma_L + \Gamma_R$, but smaller than the mechanical frequency $\omega_0$.
The corresponding hierarchy of frequencies $\Gamma \ll  T \ll \omega_0$ is obtained for example 
for the following realistic values of the parameters:
 $\Gamma = 500 \mbox{ MHz}$, $T = 50 \mbox{ mK}$ and $\omega_0/2\pi = 10 \mbox{ GHz}$.
This gives rise to the well-known Franck-Condon regime of electronic transport as studied in Ref.\cite{flensberg_tunneling_2003,mitra_phonon_2004,koch_franck-condon_2005,koch_theory_2006}.
%

%
\section{Master equation}
\label{Master_equation}
\subsection{Born-Markov approximation}
\label{Born-Markov approximation}

We define $\rho(t)$ the reduced density matrix of the mechanical oscillator and quantum dot subsystem,
obtained after tracing out the degrees of freedom of the electronic reservoirs.
In the sequential tunneling regime $(\Gamma \ll T)$, we derive a generalized master equation ruling the dynamics of the reduced density matrix
within the Born-Markov approximation \cite{schlosshauer2007decoherence,PhysRevB.83.245311,PhysRevB.80.155437,1367-2630-7-1-251}:
\begin{equation}
	 \dot{\rho}(t) = \mathcal{L} \rho(t) 
	   \label{MasterEquation1}
\end{equation}
where ${\cal L}={\cal L}_c+{\cal L}_d$ and
\begin{eqnarray}
  \mathcal{L}_c \rho &=& - i \left\lbrack \tilde{H}_0, \rho \right\rbrack
  \label{MasterEquation2} \\
  \mathcal{L}_d \rho  &=&
  \left\lbrack \mathcal{D}_h \rho - \rho \mathcal{D}_e, D^\dagger \right\rbrack
  + \mbox{H.c}
  \label{MasterEquation3}   \, .
\end{eqnarray}
The term $\mathcal{L}_c$ describes the coherent (unitary) evolution of the reduced density matrix
induced by the Hamiltonian $\tilde{H}_0$ and $\mathcal{L}_d$ describes dissipation and decoherence of the electromechanical subsystem
due to its weak coupling (tunneling term) to the electronic bath.
It involves the operator $\mathcal{D}_{\nu=e,h}$ defined as:
\begin{eqnarray}
  \mathcal{D}_{\nu} &=& \int_0^{+\infty}d\tau \mathcal{C}_{\nu}(-\tau)\mathcal{D}_{I}(-\tau)
  \label{Doperator1} \\
  \mathcal{C}_{\nu}(\tau) &=& \sum_{\alpha k} |t_{\alpha}|^2 f_{\nu\alpha}\left(\varepsilon_{\alpha k}\right) e^{i\varepsilon_{\alpha k}\tau}
  \label{Doperator2} \, ,
\end{eqnarray}
with $\mathcal{D}_{I}(-\tau)$, the operator $D$ written in interaction representation with respect
to $\tilde{H}_0$.
The correlation functions $\mathcal{C}_{\nu=e,h}(\tau)$ for the metallic reservoirs
are written in terms of the Fermi-Dirac distributions for electrons $f_{e\alpha}\left(\omega\right) \equiv f\left(\omega-\mu_{\alpha}\right)$
and holes $f_{h\alpha}\left(\omega\right) \equiv 1-f\left(\omega-\mu_{\alpha}\right)$,
with $f(\omega)=\left\lbrace e^{\beta\omega}+1 \right\rbrace^{-1}$.
The wide-band approximation enables to obtain a compact expression for the correlation
functions $\mathcal{C}_{\nu}(\omega) = \int_0^{+\infty}d\tau \mathcal{C}_{\nu}(\tau)e^{-i\left( \omega -i\eta\right)\tau}$
as:
\begin{eqnarray}
  \mathcal{C}_{\nu}(\omega) &=& \sum_{\alpha}\frac{\Gamma_{\alpha}}{2} \left\lbrace
  \frac{is_\nu}{\pi} 
  \mbox{Re }
  \Psi\left\lbrack \frac{1}{2} + \frac{i\beta}{2\pi}\left( \omega - \mu_{\alpha}\right) \right\rbrack
  + f_{\alpha \nu}\left( \omega \right)
  \right\rbrace
  \nonumber \, , \\
  \label{Correlation_Function}  
\end{eqnarray}
with $\Psi[\omega]$ the Euler digamma function \cite{abramowitz1964handbook}, obtained from the Hilbert transform of Fermi distribution
functions \cite{Bevilacqua2016} and $s_{\nu=e(h)}=1(-1)$.
The master Eq.\refe{MasterEquation1} is finally projected onto the basis
of eigenstates $\ket{q,n}$ of the Hamiltonian $\tilde{H}_0$, corresponding
to $q=0,1$ charge populating the quantum dot and $n$ vibrons populating the mechanical mode.
The eigenvalue associated to the $\ket{q,n}$ eigenstate is $\varepsilon_{qn}=q\tilde{\varepsilon}_0+n\omega_0$.
The resulting linear equations for the reduced density matrix can be solved numerically (exact Born-Markov approximation).
\subsection{Secular approximation}
\label{Secular_Approximation}

The dynamics of the coupled electromechanical system, as described by Eq.\refe{MasterEquation1}, is quite complicated.
A series of approximations can be derived in order to simplify the master equation:
(i) by first dropping in the dissipative evolution [Eq.\refe{MasterEquation3}] terms that can be incorporated into a renormalization
of $\tilde{H}_0$ (Lamb-shift terms),
(ii) second by performing a secular approximation, which enables to separate the evolution of diagonal elements of the density matrix
$\pi_{(q,n)}(t)\equiv\rho_{(q,n)(q,n)}(t)$ (populations)
from the evolution of off-diagonal terms $\sigma_{(q,n)}^{(r,m)}(t)\equiv\rho_{(q,n)(r,m)}(t)$ (coherences).
The secular approximation however, has to be done with some care, due to the equidistance between energy levels of the mechanical oscillator \cite{cohen1992atom}.
We finally obtain the following set of linear equations describing the dynamics of the damped mechanical oscillator coupled capacitively
to a quantum dot:
\begin{eqnarray}
  \dot{\pi}_{(q,n)}(t) &=& \sum_{m\in \mathbb{N}} \left\lbrace
  \Gamma^{(\bar{q},m)}_{(q,n)}\pi_{(\bar{q},m)}(t)  - 
  \Gamma^{(q,n)}_{(\bar{q},m)}\pi_{(q,n)}(t) 
  \right\rbrace 
  \label{Pauli_Master_Equation} \\
  \dot{\sigma}_{(q,n)}^{(r,m)}(t) &=& - \left\lbrack i \Omega_{(q,n)}^{(r,m)} + \frac{\Lambda_{(q,n)}^{(r,m)}}{2}\right\rbrack\sigma_{(q,n)}^{(r,m)}(t)
  \nonumber \\
  &+& \delta_{q,r}\sum_{p\in \mathbb{N}}\Xi^{(\bar{q},p)(\bar{q},p+m-n)}_{(q,n)(q,m)}\sigma_{(\bar{q},p)}^{(\bar{q},p+m-n)}(t)
  \label{Coherences_Master_Equation}  \, ,
\end{eqnarray}
with $\delta_{q,r}$ the Kronecker delta and $\bar{q}=1,0$ when $q=0,1$.
Eq.\refe{Pauli_Master_Equation} is the Pauli rate equation giving the evolution of populations.
The transition rates $\Gamma^{(q,n)}_{(\bar{q},m)}$ between the states $\ket{q,n}$ and $\ket{\bar{q},m}$
coincide with the expressions given by Fermi golden rule \cite{mitra_phonon_2004,koch_theory_2006}:
\begin{eqnarray}
  \Gamma^{(0,n)}_{(1,m)} &=& \sum_{\alpha}\Gamma_{\alpha}|Q_{n,m}|^2f_{e \alpha}(\tilde{\varepsilon}_0+(m-n)\omega_0)
  \label{Rate1} \\
  \Gamma^{(1,n)}_{(0,m)} &=& \sum_{\alpha}\Gamma_{\alpha}|Q_{n,m}|^2f_{h \alpha}(\tilde{\varepsilon}_0-(m-n)\omega_0)
  \label{Rate2} \, ,
\end{eqnarray}
with $Q_{n,m}\equiv \bra{n} Q \ket{m}$ the overlap integral between the state of the mechanical oscillator with $n$ vibrons
and the state of the displaced mechanical oscillator with $m$ vibrons \cite{mitra_phonon_2004,koch_theory_2006}.
Eq.\refe{Coherences_Master_Equation} provides the evolution of the off-diagonal elements of the density matrix.
We introduced the following quantities:
\begin{eqnarray}
  \Omega_{(q,n)}^{(r,m)} &=& \left\lbrack \left(q-r\right)\tilde{\varepsilon}_0 + \left(n-m\right)\omega_0\right\rbrack
  \label{BohrFrequencies} \\
  \Lambda_{(q,n)}^{(r,m)} &=& \sum_{p\in \mathbb{N}} \left\lbrack \Gamma^{(q,n)}_{(\bar{q},p)} + \Gamma^{(r,m)}_{(\bar{r},p)}\right\rbrack
  \label{Dissipation1} \, ,
\end{eqnarray}
with $\Omega_{(q,n)}^{(r,m)}$ the Bohr frequency associated to the states $\ket{q,n}$ and $\ket{r,m}$,
and $\Lambda_{(q,n)}^{(r,m)}$ the decay rate that is responsible for the damping of the corresponding off-diagonal element
of the density matrix.
Finally, the matrix element $\Xi^{(\bar{q},p)(\bar{q},p+m-n)}_{(q,n)(q,m)}$ is associated to the transfer of coherences between the couple of states
$\left\lbrace \ket{q,n},\ket{q,m}\right\rbrace$ and
$\left\lbrace \ket{\bar{q},p},\ket{\bar{q},p+m-n}\right\rbrace$ for the damped mechanical oscillator.
It is explicitly given by:
\begin{eqnarray}
  \Xi^{(0,p)(0,p+m-n)}_{(1,n)(1,m)} &=& \sum_{\alpha}\Gamma_{\alpha}Q_{p,n}^*Q_{p+m-n,m} f_{e\alpha}(\Omega^{0,p}_{1,n})
  \label{Dissipation2} \\
  \Xi^{(1,p)(1,p+m-n)}_{(0,n)(0,m)} &=& \sum_{\alpha}\Gamma_{\alpha}Q_{n,p}Q^*_{m,p+m-n} f_{h\alpha}(\Omega^{0,n}_{1,p})
  \label{Dissipation3} \, .
\end{eqnarray}
The evolution of the off-diagonal elements of the density matrix as described by Eqs.\refe{Coherences_Master_Equation}
was not taken into account in Refs. \cite{mitra_phonon_2004,koch_theory_2006}.
This is due to the fact that they are not needed to compute
the average electronic current in the sequential tunneling regime.
However, when dealing with the study of the mechanical oscillator dynamics, these terms are necessary.
%

\subsection{Fluctuation spectrum}
We wish now to study observable properties characterizing the dynamical state of the mechanical oscillator.
For this purpose, we will investigate the average value $\bar{A} \equiv \left\langle A \right\rangle$ as well
as the correlation function $S_{AA}(t)\equiv \left\langle \delta A(t) \delta A(0) \right\rangle$
associated to fluctuations $\delta A(t)=A(t)-\bar{A}$ of the observable
$A$ acting on the mechanical oscillator.
In the following, $A$ will stand for either the mechanical energy operator $E=\omega_0 n$ 
that is proportional
to the phonon number operator 
$n=a^\dagger a$ or for the position operator 
as defined in \Eqref{xdef}.
We further introduce the vector $\underline{\rho}(t)$ made of the matrix elements
of the reduced density matrix $\rho(t)$ (including both diagonal and off-diagonal terms).
The master Eq.\refe{MasterEquation1} can be given the compact form:
\begin{eqnarray}
  \dot{\underline{\rho}}(t) = \check{\mathcal{L}} \underline{\rho}(t)
  \label{MasterEquationSuperOperator1} \, ,
\end{eqnarray}
with $\check{\mathcal{L}}$ the super-operator associated to the linear operator $\mathcal{L}$.
Assuming a given initial condition for the density matrix $\underline{\rho}(0)$, we obtain for $\underline{\rho}(t)$:
\begin{eqnarray}
  \underline{\rho}(t) = e^{\check{\mathcal{L}} t} \underline{\rho}(0)
  \label{MasterEquationSuperOperator2} \, .
\end{eqnarray}
The stationary density matrix $\underline{\rho}^{st}$ is the solution of the equation
$\check{\mathcal{L}}\underline{\rho}^{st} = \underline{0}$, from which 
the average value of the quantum mechanical observable $A$ is obtained:
\begin{eqnarray}
  \bar{A} &=& \mbox{tr} \left(\rho^{st} A\right) \equiv\underline{w}^t \check{A}\underline{\rho}^{st}
  \label{Average1} \, ,
\end{eqnarray}
with $\underline{w}^t$ the null left-eigenvector of the $\check{\mathcal{L}}$ operator
($\underline{w}^t\check{\mathcal{L}} = \underline{0}$).
$\underline{w}^t$ applied to any vector $\underline{A}$,
reproduces the action of the quantum mechanical trace $\underline{w}^t\underline{A}=\mbox{tr}\left(A\right)$.
Defining the fluctuation spectrum of $A$ as $S_{AA}(\omega) = \int_{-\infty}^{+\infty} dt e^{i\omega t} S_{AA}(t)$ 
and using the quantum regression theorem \cite{cohen1992atom,kirton_quantum_2012}, we finally obtain:
\begin{eqnarray}
  S_{AA}(\omega) &=& - 2 \mbox{Re } \left\lbrace
  \underline{w}^t \delta \check{A} \frac{1}{ \left( i\omega - \eta \right)\check{\mbox{Id}} + \check{\mathcal{L}}} \delta\check{A} \underline{\rho}^{st}
  \right\rbrace
  \label{PDFA} \, .
\end{eqnarray}
In the following, we will consider the symmetrized fluctuation spectrum
of the $A$ operator: $S^{sym}_{AA}(\omega) = \left( S_{AA}(\omega) + S_{AA}(-\omega) \right)/2$.
%

\section{Energy fluctuation spectrum}
\label{Energy_fluctuation_spectrum}

\subsection{Dissipation of energy}
\label{Dissipation_of_energy}
%
%
\begin{figure}[tbh]
	\includegraphics[width=1.0\linewidth]{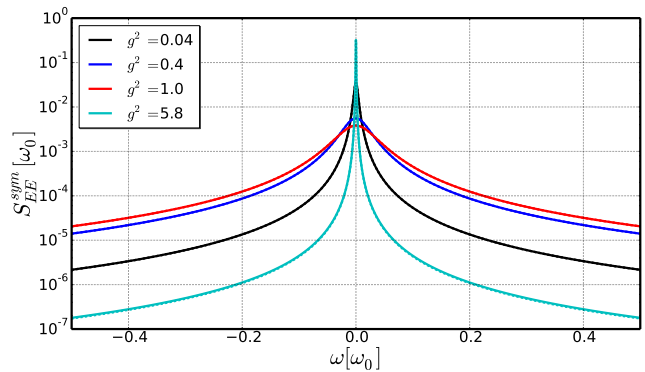}
	\caption{
		Symmetrized energy fluctuation spectrum $S^{sym}_{EE}(\omega)$ of the mechanical oscillator
		as a function of frequency $\omega$.
		Plain curves: numerical results using the secular approximation developed in Sec.~\ref{Secular_Approximation}.
		Dashed curves: analytical Lorentzian shapes provided by Eq.\refe{EnergyFluctuations2}.
		Various electromechanical coupling strengths are probed:
		$g^2=0.04,0.4,1.0,5.8$.
		Parameters common to all curves: $\Gamma=0.05\omega_0$, $\tilde{\varepsilon}_0=0$,
		$T = 0.1 \omega_0$ and $V=0.2\omega_0$.
	}
	\label{fig:S_EE}
\end{figure}
%
%
%
We first characterize the dissipation rate $\gamma_E$ of the mechanical oscillator energy.
For simplicity, we consider the regime of symmetric tunneling to the leads ($\Gamma_L=\Gamma_R=\Gamma$),
electron-hole symmetric point for the dot-level position ($\tilde{\varepsilon}_0=0$), and
symmetric bias-voltage drop ($\mu_L=-\mu_R=V/2$).
In this regime, we find that the transition rates in Eq.\refe{Rate1} and Eq.\refe{Rate2} are equal, namely
$\Gamma^{(0,n)}_{(1,m)} = \Gamma^{(1,n)}_{(0,m)}\equiv \Gamma_{n \rightarrow m}$.
This simplification enables to write a rate equation for the phonon distribution $\pi_n(t) \equiv \pi_{0,n}(t) + \pi_{1,n}(t)$ using
Eq.\refe{Pauli_Master_Equation}:
\begin{eqnarray}
  \dot{\pi}_{n}(t) &=& \sum_{m\in \mathbb{N}, m \ne n} \left\lbrace
  \Gamma_{m \rightarrow n} \pi_{m}(t) -
  \Gamma_{n \rightarrow m} \pi_{n}(t)
  \right\rbrace 
  \label{Pauli_Master_Equation2} \, .
\end{eqnarray}
In the limit of low voltage and temperature ($T,V < \omega_0$), the transition rates
simplify to:
\begin{eqnarray}
  \Gamma_{m \rightarrow n} &\approx& 2\Gamma |Q_{m,n}|^2 \theta_{m-n} + \Gamma |Q_{n,n}|^2 \delta_{n,m} 
  \label{Rate3} \, ,
\end{eqnarray}
with $\theta_{m-n}=1$ if $m>n$ and $\theta_{m-n}=0$ zero otherwise.
The meaning of Eq.\refe{Rate3} is that close to equilibrium, only transitions from higher energy states $m$ to lower energy ones $n<m$ are allowed.
The stationary phonon distribution $\pi_n^{st}$ is thus the one obtained for a mechanical oscillator in its equilibrium quantum ground state,
namely $\pi_n^{st} = \delta_{n,0}$.
\\
In order to find the energy relaxation for the mechanical oscillator,
we consider the time evolution towards steady state of a weak fluctuation $\pi_n(t) \approx \pi_n^{st} + \delta \pi_n(t)$
with $|\delta \pi_n(t)|\ll 1$.
Using Eq.\refe{Pauli_Master_Equation2} and Eq.\refe{Rate3}, the average vibron population $\bar{n}(t)=\sum_{n=1}^{+\infty} n \delta \pi_n(t)$
evolves as: 
\begin{eqnarray}
  \dot{\bar{n}}(t) &\approx& 2\Gamma \sum_{n=1}^{+\infty} n
  \sum_{m=n+1}^{+\infty} |Q_{m,n}|^2 \delta\pi_{m}(t) 
  \nonumber \\
  &-&
  2\Gamma \sum_{n=1}^{+\infty} n\sum_{m=0}^{n-1} |Q_{n,m}|^2 \delta\pi_{n}(t)
  \label{Pauli_Master_Equation4} \, ,
\end{eqnarray}
which is not a closed equation in $\bar{n}(t)$.
However, we remark that in the regime $T,V < \omega_0$, it is very unlikely that high-energy vibrational sidebands are
significantly excited.
We thus truncate the vibron distribution to the ground and first excited states
$\delta\pi_n(t) \approx \delta \pi_0(t)\delta_{n,0} + \delta \pi_1(t)\delta_{n,1}$, such that
the average vibron population becomes $\bar{n}(t)\approx \delta\pi_{1}(t)$.
This assumption is verified \textit{a posteriori} and enables to rewrite 
Eq.\refe{Pauli_Master_Equation4} in a closed form:
\begin{eqnarray}
  \dot{\bar{n}}(t) &\approx& - \gamma_E \bar{n}(t)
  \label{Pauli_Master_Equation5} \, ,
\end{eqnarray}
with:
\begin{equation}
	\gamma_E = 2\Gamma |Q_{1,0}|^2 = 2 \Gamma g^2 e^{-g^2}
  \label{EnergyDissipation2} \, .
\end{equation}
Since $\bar{E}(t)=\bar{n}(t)\omega_0$ one can identify $\gamma_E$ with the 
energy dissipation rate.
Its interpretation is straightforward.
The energy of the mechanical oscillator is damped due to the tunneling of single electrons on the dot,
which happens on a typical time scale given by the inverse electronic tunneling rate $1/\Gamma$.
The damping rate is thus proportional to $\Gamma$ and to the Franck-Condon overlap matrix element
$|Q_{01}|^2=g^2 e^{-g^2}$, which quantifies the probability of a single tunneling electron to
loose the energy of the vibrational mode and change the charge state of the dot.
Interestingly, $\gamma_E$ is a non-monotonous function of the electromechanical coupling $g$
[see Fig.~\ref{fig:FWHM_E}].
At low coupling strengths ($g < 1$), it is proportional to the square of the
electromechanical coupling $g^2$, as provided by perturbation theory.
At higher coupling strengths ($g > 1$), the damping rate decreases exponentially due to
Franck-Condon blockade: 
the charge state of the quantum dot becomes frozen thus prohibiting
dissipation to occur through charge fluctuations.
Finally, the damping rate reaches a maximum value $\gamma_E^{max}=2\Gamma/e$ for $g=1$.
\\

\subsection{Energy fluctuations}
\label{Energy_Fluctuations}
%
%
\begin{figure}[tbh]
	\includegraphics[width=1.0\linewidth]{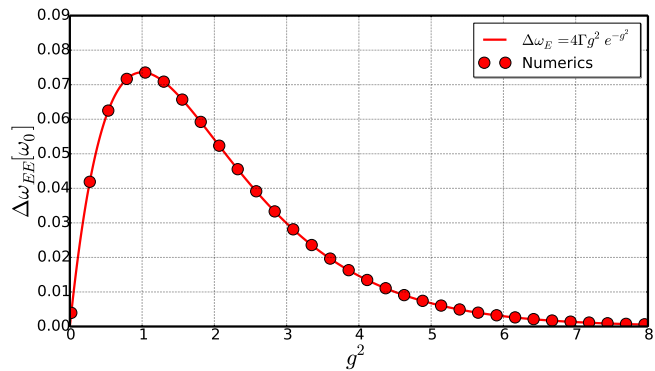}
	\caption{
		FHWM of the energy fluctuation spectrum $\Delta \omega_E$ as a function of electromechanical coupling $g^2$.
		Circles: numerical result using the secular approximation developed in Sec.~\ref{Secular_Approximation}.
		Plain curve: analytical result given by Eq.\refe{EnergyFluctuations3}.
		Parameters common to both curves: same as in Fig.~\ref{fig:S_EE}.
	}
	\label{fig:FWHM_E}
\end{figure}
%
%
%
%
We now consider energy fluctuations of the mechanical oscillator.
Consistently with the Born-Markov approximation [see Sec.~\ref{Born-Markov approximation}] and with Eq.\refe{Pauli_Master_Equation5},
the time-evolution for the mechanical energy $E(t)$ is ruled by the following Langevin equation:
\begin{eqnarray}
  \dot{E}(t) &=& - \gamma_E E(t) + \xi_E(t)
  \label{EnergyFluctuations1} \, .
\end{eqnarray}
The fluctuating part of the mechanical energy $\xi_E(t)$ is of zero average $\left\langle \xi_E(t) \right\rangle=0$
and is delta-correlated in time $\left\langle \xi_E(t)\xi_E(t')\right\rangle = D_E \delta(t-t')$.
The diffusion coefficient $D_E = 2\gamma_E \Delta n^2$ is related to
the dissipation rate $\gamma_E$ and to fluctuations of the phonon population
$\Delta n^2 = \left\langle n^2 \right\rangle-\bar{n}^2$.
At thermal equilibrium, we obtain
$D_E = 2\gamma_E n_B$, with the Bose distribution $n_B = \left\lbrace e^{\beta \omega_0}-1 \right\rbrace^{-1}$. 
After Fourier transform, Eq.\refe{EnergyFluctuations1} enables to find an analytical expression for the symmetrized spectrum $S^{sym}_{EE}(\omega)$:
\begin{eqnarray}
  S^{sym}_{EE}(\omega) &=& \frac{2\gamma_E \Delta n^2}{ \omega^2 + \gamma_E^2}
  \label{EnergyFluctuations2} \, .
\end{eqnarray}
The energy fluctuation spectrum is thus a Lorentzian centered around zero frequency with FWHM $\Delta \omega_E$ given by twice the dissipation rate:
\begin{eqnarray}
  \Delta \omega_E &=& 2 \gamma_E =  4 \Gamma g^2 e^{-g^2}
  \label{EnergyFluctuations3} \, .
\end{eqnarray}
The energy fluctuation spectrum $S^{sym}_{EE}(\omega)$ is presented in Fig.~\ref{fig:S_EE}, in the regime $T = 0.1 \omega_0$ and $V=0.2\omega_0$,
for which the mechanical oscillator is close to equilibrium.
Plain curves are computed numerically using the secular approximation developed in Sec.~\ref{Secular_Approximation}.
Dashed curves are the analytical Lorentzian shapes provided by Eq.\refe{EnergyFluctuations2}.
The extraction of the FWHM from the numerical curves is shown as a function of the electromechanical coupling $g^2$
in Fig.~\ref{fig:FWHM_E} (red circles).
The plain red curve is obtained from the analytical formula in Eq.\refe{EnergyFluctuations3}.
In both Fig.~\ref{fig:S_EE} and Fig.~\ref{fig:FWHM_E}, the perfect agreement between the numerics and the analytics,
stands for a confirmation that the broadening mechanism for energy fluctuations is indeed controlled by electronic dissipation,
so ultimately by tunneling of single electrons in and out the quantum dot.

\subsection{Bistability of the mechanical oscillator}
%
%
\begin{figure}[tbh]
	\includegraphics[width=1.0\linewidth]{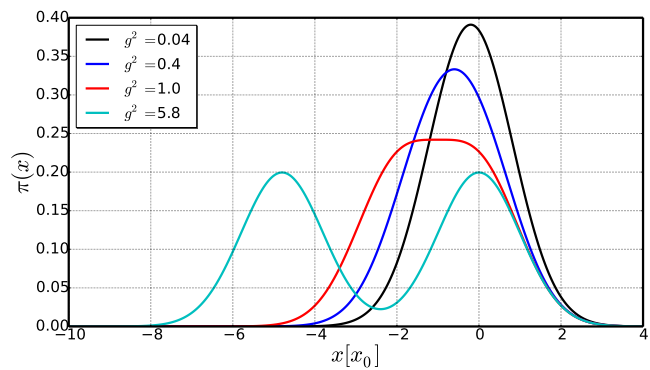}
	\caption{
		Stationary probability distribution of the oscillator position $\pi(x)$.
		Various electromechanical coupling strengths are probed:
		$g^2=0.04,0.4,1.0,5.8$.
		Parameters common to all curves: $\Gamma=0.05\omega_0$, $\tilde{\varepsilon}_0=0$,
		$T = 0.1 \omega_0$ and $V=0.2\omega_0$.
	}
	\label{fig:P_x}
\end{figure}
%
%
%
%
%
%
\begin{figure}[tbp]
  \includegraphics[width=1.0\linewidth]{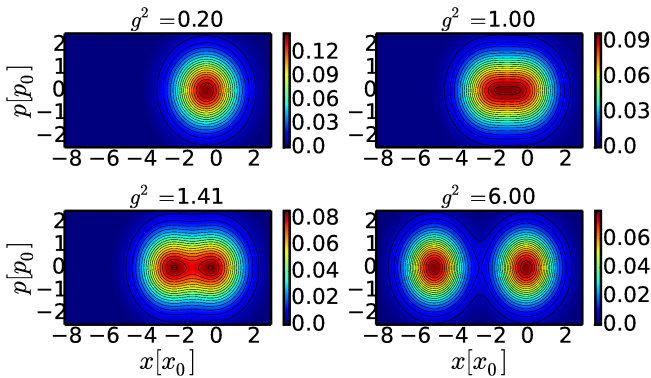}
  \caption{
    Wigner distribution $W(x,p)$ for the mechanical oscillator as a function of the oscillator position $x$
    and momentum $p$.
    2D-maps obtained for various values of the electromechanical coupling:
    $g^2 = 0.2,1.0,\sqrt{2},6.0$.
    Parameters common to all panels: $\Gamma=0.05\omega_0$, $\tilde{\varepsilon}_0=0$, $T = 0.1 \omega_0$, and $V=0.2\omega_0$.
  }
\label{fig:Wigner}
\end{figure}
%
%
In this section, we compute the stationary probability distribution $\pi(x)$ of the mechanical oscillator position.
The stationary density matrix of the mechanical oscillator coupled to the quantum dot is
approximatively diagonal in the basis of the eigenstates $\ket{q n}$, namely $\rho^{st}\approx\sum_{q,n} \pi_{(q,n)} \ket{q n}\bra{q n}$.
The stationary distribution $\pi(x)$ is thus approximated by:
\begin{eqnarray}
  \pi(x) &\approx& \sum_{n\in \mathbb{N}} \left\lbrace
  \pi_{(0n)}|\phi_n(x)|^2 +
  \pi_{(1n)}|\phi_n(x+2gx_0)|^2
  \right\rbrace
  \nonumber \, .\\
  \label{Px1} 
\end{eqnarray}
In Eq.\refe{Px1}, $\phi_n( x )$ is the wave-function of the mechanical oscillator's $n$-th eigenstate:
\begin{eqnarray}
  \phi_n( x ) &=& \frac{\left( 2\pi \right)^{-\frac{1}{4}}}{\sqrt{x_0 2^n n!}}H_n\left[ \frac{x}{x_0\sqrt{2}} \right] \exp{\left\lbrack 
    -\left( \frac{x}{2x_0} \right)^2 \right\rbrack}
  \label{Px2} \, ,
\end{eqnarray}
with $H_n\left[ x \right]$ the $n$-th Hermite polynomial \cite{cohenquantummechanics}. 
We present in Fig.~\ref{fig:P_x}, the probability distribution $\pi(x)$ obtained with the same parameters as in
Sec.~\ref{Dissipation_of_energy}, for which the mechanical oscillator is close to its quantum ground state $n=0$.
We find that at low electromechanical coupling ($g<1$), the probability distribution $\pi(x)$ has a single peak, and the
mechanical oscillator is monostable.
At larger couplings ($g>1$), the distribution develops two peaks, and the
mechanical oscillator becomes bistable. 
The transition between the monostable behavior and the bistable one happens for $g=1$, for which
the distribution has a very flat top.
The mechanism responsible for this transition, is the following. 
For any value of the coupling strength $g$, the mechanical oscillator has two stable equilibrium positions located at
$x = 0$ and $x = -2g x_0$, for which the charge state of the dot is respectively frozen
at $q=0$ and $q=1$.
The double peak structure is resolved whenever the average shift of the equilibrium position  $\overline{\Delta x}=-2gx_0\left\langle q \right\rangle \equiv - g x_0$ induced by electromechanical coupling
overcomes the zero-point quantum fluctuations $-\overline{\Delta x} = g x_0 > x_0$.
It is interesting to notice that the transition point ($g=1$) coincides with the value of the
electromechanical coupling for which the damping of the mechanical oscillator is maximum
[see Fig.~\ref{fig:FWHM_E} in Sec.~\ref{Energy_Fluctuations}].
We complete the picture of the transition to bistability by showing in Fig.~\ref{fig:Wigner} the 2D-plots representing the
mechanical oscillator Wigner distribution \cite{PhysRev.40.749,LEE1995147} defined as
$W(x,p)=\frac{1}{2\pi}\int dy \bra{x+\frac{y}{2}}\rho\ket{x-\frac{y}{2}}e^{-ipy}$,
with $p$ the oscillator momentum expressed in units
of $p_0=\sqrt{2m\omega_0}$.
We find that the Wigner distribution goes smoothly from a single-peak distribution at low
electromechanical coupling $g^2=0.2$ towards a double-peak distribution at higher-coupling $g^2=6.0$.
The critical coupling $g^2=1$ is characterized by a flattened distribution, in agreement with Fig.~\ref{fig:P_x}.
It is to be noted that no negative contribution to the Wigner distribution is obtained.
This is due to the fact that the Wigner distribution of a harmonic oscillator in its quantum ground state
is a Gaussian positive distribution \cite{LEE1995147}.
%
%
\section{Displacement fluctuation spectrum}
\label{Displacement_fluctuation_spectrum}
\subsection{Oscillator decoherence time}
\label{Oscillator_decoherence_time}
In this section, we investigate the evolution of the average of the $X$-operator:
$\overline{X}(t)=x_0\left\lbrace \overline{a(t)} + \overline{a^\dagger(t)} \right\rbrace$, 
obtained as:
\begin{eqnarray}
  \overline{X}(t) = 2 x_0 \sum_{n=0}^{+\infty} \sqrt{n+1} \mbox{ Re}\left\lbrace \rho^{(mec)}_{n n+1}(t)\right\rbrace
  \, , \label{X_average} 
\end{eqnarray}
with $\rho^{(mec)}_{nm}(t) = \sum_{q=0,1} \rho_{(qn)(qm)}(t)$ the reduced density matrix of the mechanical oscillator, obtained
after tracing out the charge degrees of freedom of the dot.
Note that the physical displacement is given by \Eqref{xdef} and implies also the charge operator $n_d$.
We will see that the relevant fluctuations of $n_d$ are at low-frequency, allowing
to regard $x \approx X$ at high-frequency $\omega\approx\omega_0$.  
We consider the same regime of low voltage and temperature ($T, V < \omega_0$) and symmetric electron-hole point ($\tilde{\varepsilon}_0=0$), as in Sec.~\ref{Dissipation_of_energy}.
Within the same approximation consisting of truncating the oscillator reduced density matrix to at most one vibron excitation ($n,m=0,1$), the
average position is obtained as $\overline{X}(t) \approx 2 x_0 \mbox{ Re} \left\lbrace \rho^{(mec)}_{01}(t) \right\rbrace$.
Using Eq.\refe{Coherences_Master_Equation} and Eq.\refe{Rate3}, one can show after some algebra, that in this
quasi-equilibrium regime, the time evolution of $\rho^{(q)}_{01}(t)\equiv \rho_{(q0)(q1)}(t)$ is given by:
\begin{eqnarray}
  \dot{\rho}^{(q)}_{01}(t) &\approx&
  \left\lbrace i\omega_0 - \Gamma \left\lbrack
  |Q_{10}|^2 + \frac{|Q_{00}|^2 + |Q_{11}|^2}{2}
  \right\rbrack\right\rbrace \rho^{(q)}_{01}(t) 
  \nonumber \\
  &+& \Gamma Q_{00} Q_{11} \rho^{(\bar{q})}_{01}(t) \, ,
  \label{Coherences_Master_Equation2}  
\end{eqnarray}
The first term in Eq.\refe{Coherences_Master_Equation2} describes the coherent evolution
between the states of same charge $q=0,1$ and different number of phonons $n=0$ and $m=1$.
The second (third) term describes the incoherent
evolution between the states of same (different) charge $q=0,1$ ($\bar{q}=1,0$)
and different number of phonons $n=0$ and $m=1$,
due to electromechanical coupling.
We deduce from Eq.\refe{Coherences_Master_Equation2} the evolution of the oscillator reduced density matrix:
\begin{eqnarray}
  &&\dot{\rho}^{(mec)}_{01}(t) \approx 
  \left\lbrace i\omega_0 - \gamma_X \right\rbrace \rho^{(mec)}_{01}(t)
  \label{Coherences_Master_Equation3} \\
  &&\gamma_X = \Gamma \left\lbrace
  |Q_{10}|^2 + \frac{|Q_{00}|^2 + |Q_{11}|^2}{2} - Q_{00}Q_{11}
  \right\rbrace \, ,
  \label{Decoherence_Rate3}
\end{eqnarray}
with $\gamma_X$ the decoherence rate of the  mechanical oscillator.
Eq.\refe{Coherences_Master_Equation3} enables to
write the equation for $\overline{X}(t)$:
\begin{eqnarray}
  &&\ddot{\overline{X}}(t) + 2\gamma_X\dot{\overline{X}}(t) + \left( \omega_0^2 + \gamma_X^2 \right)\overline{X}(t) = 0
  \label{Evolution_X1} \, , \\
  &&\gamma_X = \Gamma g^2 \left\lbrack 1 + \frac{g^2}{2} \right\rbrack e^{-g^2} \, .
  \label{Evolution_X2}
\end{eqnarray}
Eq.\refe{Evolution_X1} coincides with the equation of motion of a classical damped harmonic oscillator.
Interestingly, the decoherence rate $\gamma_X$ as given by Eq.\refe{Evolution_X2}, 
does not coincide with the energy dissipation rate $\gamma_E/2$ obtained in Eq.\refe{EnergyDissipation2}.
The decoherence rate can be also written as:
\begin{eqnarray}
  \gamma_X &=& \frac{\gamma_E}{2} + \gamma_\phi
  \label{Decoherence_Rate1} \\
  \gamma_\phi &=& g^2 \frac{\gamma_E}{4} = \frac{\Gamma}{2} g^4 e^{-g^2}
  \label{Decoherence_Rate2} \, .
\end{eqnarray}
The first term $\gamma_E/2$ in Eq.\refe{Decoherence_Rate1} gives the standard contribution of the dissipation
to the decoherence of the mechanical oscillator.
The second term $\gamma_\phi$ is an additional dephasing rate.
This term has some interesting consequences.
First of all, the decoherence rate $\gamma_X$ of the mechanical oscillator is larger than the contribution induced by pure energy dissipation:
$\gamma_X \ge \gamma_E/2$.
Then, $\gamma_X$ as a function of $g^2$ reaches a maximum for a value of the electromechanical coupling $g^2=\sqrt{2}$ that is larger
than the value $g^2=1$ for which dissipation is maximal [see Fig.~\ref{fig:FWHM_x}].
In other words, the maximal decoherence rate is obtained after entering in the region of bistability of the mechanical oscillator, while the
maximal dissipation rate coincides with the frontier between the monostable and bistable region
[see Fig.~\ref{fig:P_x} and \ref{fig:FWHM_E}].

\subsection{Microscopic mechanism for decoherence}
\label{Decoherence_Mechanism}
%
\begin{figure}[tbh]
  \includegraphics[width=1.0\linewidth]{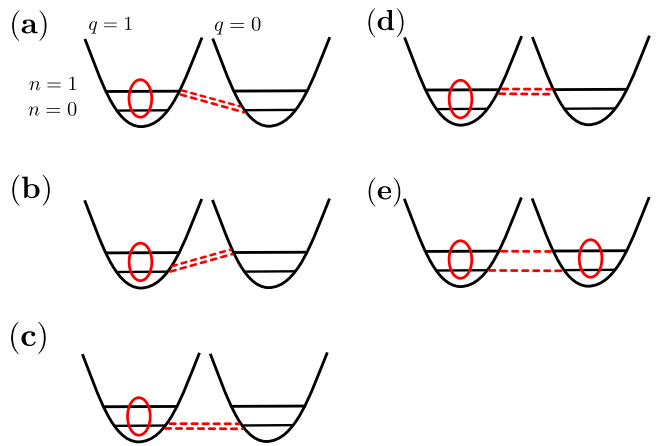}
  \caption{
    Schematics of the microscopic processes responsible for the decoherence rate $\gamma_X$ of the off-diagonal element of the mechanical oscillator
    density-matrix $\rho^{(mec)}_{01}(t)=\sum_{q=0,1}\rho^{(q)}_{01}(t)$.
    $(a)$ and $(b)$: Inelastic processes (red dashed) responsible for energy dissipation $\gamma_E$. One mechanical vibron is absorbed while the charge state
    of the quantum dot is modified.
    $(c)$, $(d)$, $(e)$ and $(f)$: Elastic processes (red dashed) responsible for dephasing $\gamma_\phi$. No mechanical vibron is emitted nor absorbed, while the charge state
    of the quantum dot is modified.
    $(e)$: Transfer of coherences (red dashed). 
    Common to all: The red circles in the charge sector $q=0,1$ stands for the matrix element $\rho^{(q)}_{01}(t)$ in Eq.\refe{Coherences_Master_Equation2}.
    It is coupled either to itself or to the matrix element $\rho^{(\bar{q})}_{01}(t)$ of the complementary charge sector $\bar{q}=1,0$. 
  }
\label{fig:Processus_Decoherence}
\end{figure}
%
%
%
The decoherence rate is obtained by the additive contribution of several elementary microscopic processes
in Eq.\refe{Decoherence_Rate3}.
The first term $\propto \Gamma|Q_{10}|^2$ is the degenerate contribution of the processes picture in
Fig.~\ref{fig:Processus_Decoherence}-(a) and (b).
Those processes, responsible for energy dissipation $\gamma_E$, are
inelastic processes during which one mechanical vibron is absorbed, while the charge state
of the quantum dot is modified.
The second and third terms $\propto \Gamma/2 \left( |Q_{00}|^2 + |Q_{11}|^2 \right)$ are purely elastic processes
for which no mechanical vibron is emitted nor absorbed, while the charge state of the quantum dot is modified.
They are presented in Fig.~\ref{fig:Processus_Decoherence}-(c) and (d), respectively.
The last terms $\propto -\Gamma Q_{00}Q_{11}$ are elastic processes corresponding
to transfer of coherences between pair of states $(00),(01)$ and $(10),(11)$.
They are pictured in Fig.~\ref{fig:Processus_Decoherence}-(e).
It is interesting to notice that the dephasing rate $\gamma_\phi$ in Eq.\refe{Evolution_X2} originates
entirely from the elastic processes.
Those are higher-order terms in the electromechanical coupling.
Note that the standard description of a quantum damped harmonic oscillator \cite{schlosshauer2007decoherence} does not 
predicts a difference between the decoherence rate and half the dissipation rate.
This originates here from the presence of the additional charge degree of freedom.

\subsection{Displacement fluctuations}
%
%
\begin{figure}[tbh]
  \includegraphics[width=1.0\linewidth]{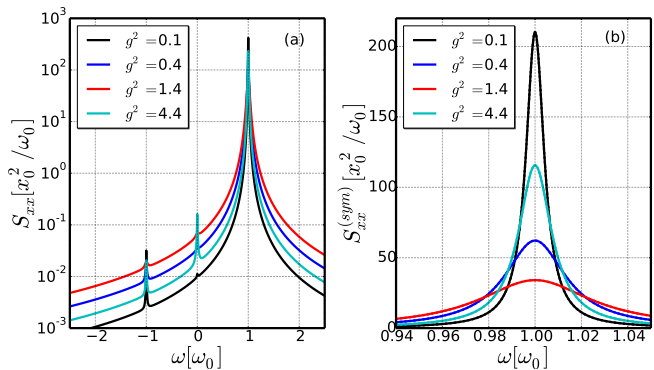}
  \caption{
    Fluctuation spectrum of the oscillator displacement $S_{xx}(\omega)$ as a function of frequency $\omega$.
    (a) Asymmetric spectrum $S_{xx}(\omega)$ computed within the secular approximation developed in Sec.~\ref{Secular_Approximation}.
    Various electromechanical coupling strengths are probed:
    $g^2=0.1,0.4,1.4,4.4$.
    (b) Symmetrized displacement spectrum $S^{sym}_{xx}(\omega)$ around the phonon-emission peak at $\omega \approx \omega_0$.
    Plain curves: numerical results.
    Dashed curves: analytical Lorentzian shapes provided by Eq.\refe{S_XX_1}.
    Parameters common to both panels: $\Gamma=0.05\omega_0$, $\tilde{\varepsilon}_0=0$, $T = 0.1 \omega_0$ and $V=0.2\omega_0$.
  }
\label{fig:S_xx_1}
\end{figure}
%
%
%
We consider now the influence of the stochastic force fluctuations acting on $X$: $\xi_X(t)$.
The corresponding Langevin equation reads:
\begin{eqnarray}
  &&\ddot{X}(t) + 2\gamma_X\dot{X}(t) + \left( \omega_0^2 + \gamma_X^2 \right) X(t) = \xi_X(t) \, ,
  \label{Langevin_x1} 
\end{eqnarray}
with $\left\langle \xi_X(t)\xi_X(t')=D_{X}\delta(t-t')\right\rangle$.
The diffusion coefficient for the fluctuations of $X$ is defined as $D_X$ and can in general 
be expressed in terms of its variance 
$D_{X}=4\omega_0^2\gamma_X \Delta X^2$, with
$\Delta X^2=\left\langle X^2 \right\rangle - \overline{X}^2$.
At thermal equilibrium, the fluctuation-dissipation theorem \cite{PhysRev.83.34} gives
$D_X = 4\omega_0^2x_0^2 \gamma_X\coth{ \left(\beta\omega_0/2 \right)}$.
After Fourier transforming Eq.\refe{Langevin_x1}, we obtain in the limit of weak electronic
damping ($\gamma_X \ll \omega_0$):
\begin{eqnarray}
  S^{sym}_{XX}(\omega) &\approx& \sum_{s=\pm 1}\frac{\gamma_X \Delta X^2}{ \left( \omega + s\omega_0 \right)^2 + \gamma_X^2}
  \label{S_XX_1} \, .
\end{eqnarray}
The symmetrized $X$ fluctuation spectrum is thus a sum of Lorentzians centered at frequencies $\omega = \pm \omega_0$.
Its FWHM $\Delta \omega_X$ is given by:
\begin{eqnarray}
  \Delta \omega_X &=& \frac{\Delta \omega_E}{2} + \Delta \omega_\phi = 2\Gamma g^2 \left\lbrack 1 + \frac{g^2}{2} \right\rbrack e^{-g^2}
  \label{FWHM_X_1} \, ,
\end{eqnarray}
with the contribution of dephasing $\Delta \omega_\phi=2\gamma_\phi$.
The displacement fluctuation spectrum for the oscillator position $x = X - 2gx_0n_d$ reads:
\begin{eqnarray}
  S_{xx}(\omega) &=& S_{XX}(\omega) + 4g^2x_0^2S_{n_dn_d}(\omega)
  \label{SXX_2} \\
  &-& 2gx_0 \left\lbrace S_{Xn_d}(\omega) + S_{n_dX}(\omega) \right\rbrace \,.
\end{eqnarray}
It is the sum of three terms: (i) the contribution of thermomechanical noise $S_{XX}(\omega)$,
(ii) a contribution of charge noise $S_{n_dn_d}(\omega)$ shifting randomly the mechanical oscillator equilibrium position,
(iii) a contribution associated to correlations between the charge state of the dot and the oscillator
position $S_{Xn_d}(\omega) + S_{n_dX}(\omega)$.
The symmetrized charge noise contribution can be evaluated with the same methods as derived in Sec.~\ref{Dissipation_of_energy}.
We obtain for the total symmetrized displacement spectrum:
\begin{eqnarray}
  S_{xx}^{sym}(\omega) &\approx& \sum_{s=\pm 1}\frac{\gamma_X \Delta X^2}{ \left( \omega + s\omega_0 \right)^2 + \gamma_X^2} + 2g^2x_0^2\frac{\gamma_E}{ \omega^2 + \gamma_E^2}
  \nonumber  \, ,\\
  \label{SXX_3} 
\end{eqnarray}
where we neglected the mixed therms $X n_d$ since the two quantities fluctuate at very different frequency scales: $n_d$ at low frequencies $\omega < \gamma_E \ll \omega_0$, 
and $X$ at $|\omega-\omega_0|\ll \gamma_X$.
Fig.~\ref{fig:S_xx_1}-(a) shows the displacement spectrum $S_{xx}(\omega)$ of the mechanical oscillator as a function of frequency,
computed numerically within the secular approximation.
The spectrum of this quantum noise is strongly asymmetric.
It has a main peak at $\omega\approx \omega_0$ associated to phonon emission, which dominates the spectrum
at low temperature and voltage (only phonon emission is possible at low temperature).
A secondary peak is observed at $\omega\approx -\omega_0$ associated to phonon absorption.
Its height is very weak since phonon absorption is strongly suppressed for a mechanical oscillator close to its
quantum mechanical ground state.
Finally a last peak is observed at low frequencies $\omega\approx 0$, associated to the contribution of charge noise in Eq.\refe{SXX_2}.
The symmetrized noise $S^{sym}_{xx}(\omega)$ is presented in Fig.~\ref{fig:S_xx_1}-(b) close to the phonon emission peak.
The analytical curves (dashed curves) obtained with Eq.\refe{S_XX_1} are perfectly matching the
curves computed numerically (plain curves).
%

%
%
%
\begin{figure}[tbp]
  \includegraphics[width=1.0\linewidth]{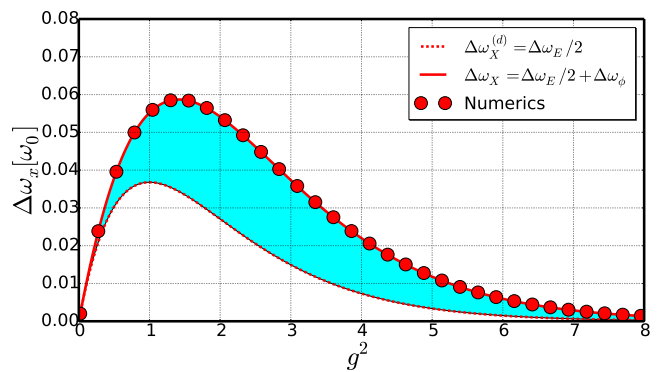}
  \caption{
    FHWM of the displacement fluctuation spectrum $\Delta \omega_x$ as a function of electromechanical coupling $g^2$.
    Circles: numerical result using the secular approximation developed in Sec.~\ref{Secular_Approximation}.
    Dashed curve: analytical result for the contribution induced by dissipation $\Delta \omega^{(d)}_X = \Delta\omega_E/2$.
    Plain curve: analytical result including the additional contribution of dephasing $\Delta\omega_\phi$ (filled blue sector)
    as given by Eq.\refe{FWHM_X_1}.
    Parameters common to both curves: same as in Fig.~\ref{fig:S_xx_1}.
  }
\label{fig:FWHM_x}
\end{figure}
%
%
The dependence of the FWHM $\Delta\omega_x$ as a function of electromechanical coupling $g^2$ is shown
in Fig.~\ref{fig:FWHM_x}.
Here also, the agreement between the analytical formula in Eq.\refe{FWHM_X_1} (plain curve) and the numerics
(circles) is very good.
This validates the scenario of decoherence presented in Sec.~\ref{Decoherence_Mechanism}, that results
from the combination of dissipation due to inelastic processes and dephasing induced by elastic processes.
%

\section{Voltage dependence}
\label{Voltage_dependence}

\subsection{Heating of the mechanical oscillator}
\label{Heating_mechanical_oscillator}
%
%
%
\begin{figure}[tbh]
	\includegraphics[width=1.0\linewidth]{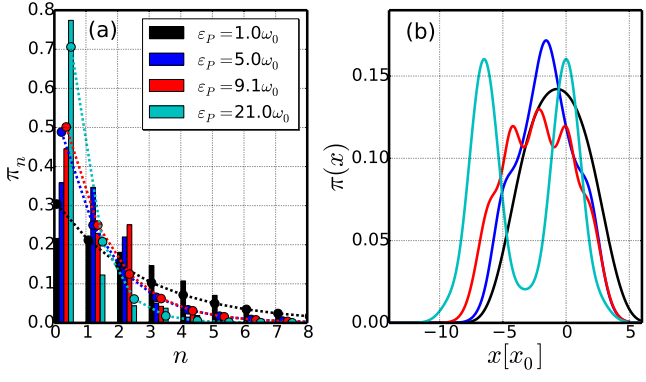}
	\caption{
		(a) Stationary distribution of the vibronic population $\pi_n$.
		Histograms obtained for various values of the electromechanical coupling:
		$\epsilon_P[\omega_0] = 1.0,5.0,9.1,21.0$.
		Corresponding average phonon population: $\bar{n} = 2.29,1.05,0.99,0.42$
		and effective temperature: $T_{eff}[\omega_0] = 2.76,1.49,1.44,0.82$.
		Circle-Dashed curves: thermal distributions $\pi^{th}_n$ with effective
		temperature $T_{eff}$ having the same average phonon number $\bar{n}$.
		(b) Corresponding stationary probability distribution of the oscillator position $\pi(x)$.
		Parameters common to both panels: $\Gamma=0.05\omega_0$, $\tilde{\varepsilon}_0=0$, $T = 0.1 \omega_0$, and $V=4.5\omega_0$.
	}
	\label{fig:P_x_Voltage}
\end{figure}
\begin{figure}[tbp]
  \includegraphics[width=1.0\linewidth]{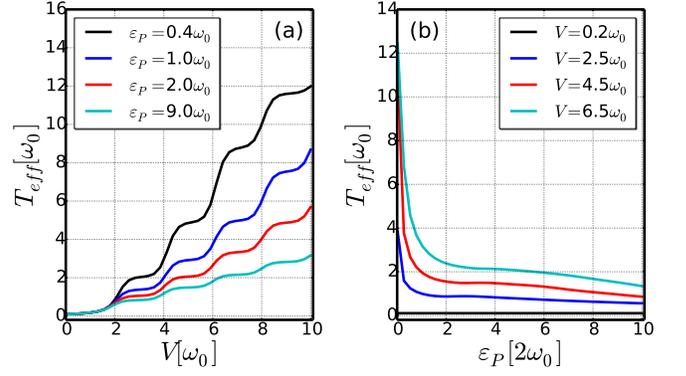}
  \caption{
    (a) Effective temperature $T_{eff}[\omega_0]$ of the mechanical oscillator as a function of bias-voltage $V$,
    for various values of the electromechanical coupling: $\epsilon_P[\omega_0] = 0.4,1.0,2.0,9.0$.
    (b) Same plot as a function of electromechanical coupling $\epsilon_P/2$,
    for various values of the bias-voltage: $V[\omega_0] = 0.2,2.5,4.5,6.5$.
    Parameters common to both panels: $\Gamma=0.05\omega_0$, $\tilde{\varepsilon}_0=0$, $T = 0.1 \omega_0$.
  }
\label{fig:T_eff}
\end{figure}

%
%
%
%
In Sec.~\ref{Displacement_fluctuation_spectrum}, we studied the dynamical properties of the mechanical oscillator
at low voltages and temperatures ($T,V < \omega_0$).
In this section, we will unravel the effect of imposing a bias-voltage larger than the typical vibron frequency
$V/2 > \omega_0$, keeping the temperature of the electronic environment 
at low values $T \ll  \omega_0$.
The main physical consequence of increasing the bias-voltage is to open an additional 
inelastic channel each time the bias-voltage
crosses a multiple of the vibron frequency $V/2 > n\omega_0$, thus modifying the expression for the transition
rates in Eq.\refe{Rate3} to
\begin{eqnarray}
  \Gamma_{m \rightarrow n} &\approx& \Gamma \sum_{\alpha=\pm} |Q_{m,n}|^2 \theta\left\lbrack \alpha \frac{V}{2} - \left( n - m \right)\omega_0 \right\rbrack  
  \label{Rate4} \, .
\end{eqnarray}
This gives rise to new possibilities of exciting vibrons in the rate equation Eq.\refe{Pauli_Master_Equation2} and thus to heat up
the mechanical oscillator. 
\\
We show in Fig.~\ref{fig:P_x_Voltage}-(a) the stationary out-of-equilibrium phonon distribution $\pi_n$ under a
bias-voltage $V=4.5\omega_0$.
In contrast to Sec.~\ref{Displacement_fluctuation_spectrum}, where only the ground state of the mechanical oscillator was significantly populated,
the phonon distribution now spreads up to high-energy excited vibronic states.
In the regime we investigate, this spreading is interpreted as a bias-induced heating of the mechanical oscillator.
In order to quantify it more precisely, we compared the phonon distribution $\pi_n$ (histograms in Fig.~\ref{fig:P_x_Voltage}-(a)) computed numerically to an effective
thermal distribution $\pi^{th}_n$ (circle-dashed curves in Fig.~\ref{fig:P_x_Voltage}-(a)) defined as
\begin{eqnarray}
  \pi^{th}_n &=& \left( 1 - e^{-\beta_{eff}\omega_0} \right) e^{-n\beta_{eff}\omega_0}
  \label{thermal} \, , \\
  \beta_{eff} &\equiv& \frac{1}{T_{eff}} = \frac{1}{\omega_0} \ln\left( \frac{ 1 + \bar{n} }{\bar{n}} \right)
  \label{Temperature_eff} \, .
\end{eqnarray}
The effective temperature $T_{eff}$ in Eq.\refe{Temperature_eff} is chosen in such a way to reproduce the exact
average vibron population $\bar{n}$ computed from the distribution $\pi_n$.
We find that for various electromechanical couplings $g^2=\epsilon_P/2\omega_0$, the vibron distribution $\pi_n$ is not far from the fitted thermal distribution
$\pi^{th}_n$ of Eq.\refe{thermal}.
%
%
At low $\epsilon_P=\omega_0$, the mechanical oscillator is heated above the temperature of the electronic environment $T_{eff} \approx 2.76\omega_0 \gg T = 0.1 \omega_0$.
Upon increasing the electromechanical coupling to $\epsilon_P=21.0\omega_0$, 
the effective temperature decreases down to $T_{eff}\approx 0.82\omega_0$.
The obtained effective temperature depends on both voltage $V$ and electromechanical coupling $\epsilon_P$ \cite{pistolesi_cooling_2009,traverso_ziani_electrical_2011,PhysRevB.83.245311,armour_classical_2004},
as shown
in Fig.~\ref{fig:T_eff}-(a) and Fig.~\ref{fig:T_eff}-(b).
We find that at voltages much lower than the vibron frequency ($V/2 \ll \omega_0$), the effective temperature
converges to the environment temperature $T_{eff}\approx 0.1\omega_0$, independently of the coupling strength,
as expected for a mechanical oscillator at thermal equilibrium.
Upon increasing the bias-voltage with $V/2 > \omega_0$, the effective temperature $T_{eff}$ becomes larger than $T$ \cite{armour_classical_2004},
consistently with Fig.~\ref{fig:P_x_Voltage}-(a).
The main tendency is a step-wise increase of $T_{eff}$ each time a vibronic sideband is excited.
At sufficiently high voltage, the step-wise increase of $T_{eff}$ becomes in average linear in $V$ with a slope that increases with decreasing $\epsilon_P$:
the smaller the electromechanical coupling, the higher the effective temperature \cite{PhysRevB.83.125419}.
%

%
%
%
%
%
%
Finally, we plot in Fig.~\ref{fig:P_x_Voltage}-(b) the stationary probability distribution of the oscillator position $\pi(x)$, for
the same range of parameters as in Fig.~\ref{fig:P_x_Voltage}-(a).
We find that, similarly to the quasi-equilibrium case [see Fig.~\ref{fig:P_x}], $\pi(x)$ undergoes a transition from a monostable situation (one peak)
at low coupling $\epsilon_P = \omega_0$ to a 
bistable situation (two peaks) at sufficiently high-coupling strength $\epsilon_P=21.0\omega_0$.
However, in contrast to Fig.~\ref{fig:P_x}, the intermediate regime ($\epsilon_P \approx 9.1\omega_0$) is characterized by a multistable situation for which
the distribution $\pi(x)$ develops two minima rather than a single broad maximum.
This difference is due to the fact that in this regime $\bar{n}\approx 1$, so that not only the ground state of the mechanical oscillator
($n=0$) contributes significantly to Eq.\refe{Px1} but also the first excited states ($n=1,2$).
%

\subsection{Displacement fluctuation spectrum}
%
%
\begin{figure}[tbh]
  \includegraphics[width=1.0\linewidth]{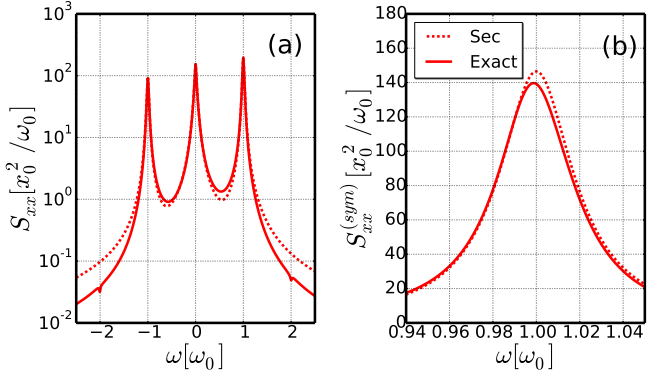}
  \caption{
    Fluctuation spectrum of the oscillator displacement $S_{xx}(\omega)$ as a function of frequency $\omega$.
    (a) Asymmetric spectrum $S_{xx}(\omega)$ computed numerically.
    Plain curve: full Born-Markov result as developed in Sec.~\ref{Born-Markov approximation}.
    Dashed curve: secular approximation developed in Sec.~\ref{Secular_Approximation}.
    (b) Same curves for the symmetrized displacement spectrum $S^{sym}_{xx}(\omega)$ computed around the phonon-emission peak at $\omega \approx \omega_0$.
    Parameters common to both panels: $\Gamma=0.05\omega_0$, $\tilde{\varepsilon}_0=0$, $T = 0.1 \omega_0$, $\epsilon_P=5.0\omega_0$ and $V=4.5\omega_0$.
  }
\label{fig:S_xx_Full_1}
\end{figure}
%
%
%
In this section, we investigate the role of the bias-voltage on the displacement fluctuation spectrum $S_{xx}(\omega)$.
In contrast to Sec.~\ref{Displacement_fluctuation_spectrum}, it is more difficult to obtain analytical insight on
the $S_{xx}(\omega)$ curves.
This is due to heating of vibron excitations, that precludes a simple truncation of the master equation
[see Eq.\refe{Pauli_Master_Equation} and Eq.\refe{Coherences_Master_Equation}] for the vibron mode.
One can consider the limit of vanishing damping and decoherence
rates $\gamma_E, \gamma_X \rightarrow 0$. 
In this limit, we compute the correlation function $\left\langle \delta A(t) \delta A(0) \right\rangle$ of any operator $A$
taking into account only the coherent evolution with respect to the free Hamiltonian $\tilde{H}_0$ in
Eq.\refe{MasterEquation1}.
Similarly to Eq.\refe{SXX_3}, the displacement fluctuation spectrum $S_{xx}(\omega)$ can be approximated as the sum of a thermomechanical noise $S_{XX}(\omega)$
plus a contribution due to low-frequency charge noise fluctuations of the dot
\begin{eqnarray}
  S_{XX}(\omega) &=& S_{abs}(\omega) + S_{em}(\omega) + 2\pi g^2 x_0^2 \delta \left( \omega \right) 
  \label{SXX_Nodamping} \, ,\\
  S_{abs}(\omega) &\approx& 2\pi x_0^2 \bar{n} \delta \left( \omega + \omega_0 \right)
  \label{Sabs_Nodamping} \, ,\\
  S_{em}(\omega) &\approx& 2\pi x_0^2 \left( 1 + \bar{n} \right) \delta \left( \omega - \omega_0 \right) 
  \label{Sem_Nodamping} \, . 
\end{eqnarray}
The thermomechanical noise spectrum in Eq.\refe{SXX_Nodamping}
is composed of an absorption noise $S_{abs}(\omega)$ of height proportional to the (voltage and coupling-dependent) average phonon population $\bar{n}$
plus an emission noise $S_{em}(\omega)$ of height proportional to $1+\bar{n}$.
The ratio between the emission noise and absorption noise $S_{em}(\omega)/S_{abs}(\omega)$
is proportional to $\left( 1 + \bar{n} \right)/\bar{n} = e^{\beta_{eff}\omega_0}$, and is thus
related the oscillator effective temperature $T_{eff}$ [see Eq.\refe{Temperature_eff}].
The symmetrized thermomechanical noise is readily obtained as
\begin{eqnarray}
  S^{sym}_{XX}(\omega) &\approx& 2\pi x_0^2 \left( \bar{n} + \frac{1}{2}\right)
  \sum_{s=\pm} \delta \left( \omega + s\omega_0 \right) 
  \label{Ssym_Nodamping}  \, .
\end{eqnarray}
$S^{sym}_{XX}(\omega)$ has thus a height proportional to the oscillator average mechanical energy.
Interestingly, Eq.\refe{Ssym_Nodamping} recovers the limits $\gamma_E,\gamma_X \rightarrow 0$ in Eq.\refe{S_XX_1}, obtained for the
case of an oscillator in the low bias and temperature regimes.
%

%
%
\begin{figure}[tbp]
  \includegraphics[width=1.0\linewidth]{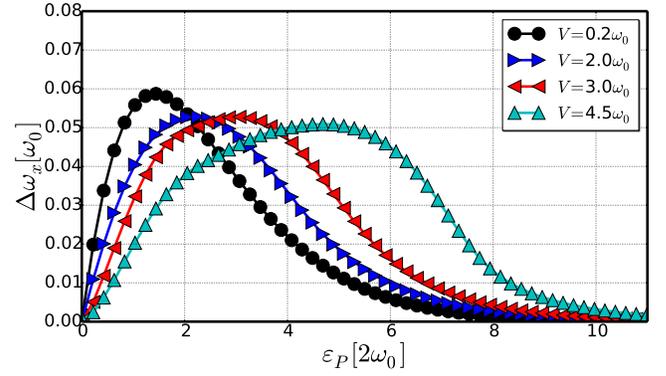}
  \caption{
    FHWM of the displacement fluctuation spectrum $\Delta \omega_x$ as a function of electromechanical coupling $\epsilon_P/2$.
    Numerical results using the secular approximation developed in Sec.~\ref{Secular_Approximation}.
    Various voltage biases are probed:
    $V[\omega_0] = 0.2,2.0,3.0,4.5$.
    Parameters common to all curves, apart from voltage: same as in Fig.~\ref{fig:S_xx_1}.
  }
\label{fig:FWHM_x_Voltage}
\end{figure}
%
%
%
We present in Fig.~\ref{fig:S_xx_Full_1}-(a) the displacement fluctuation spectrum computed numerically, using either the full Born-Markov result (plain curve)
as developed in Sec.~\ref{Born-Markov approximation} or the secular approximation (dashed curve) developed in Sec.~\ref{Secular_Approximation}.
In contrast to Fig.~\ref{fig:S_xx_1}, the spectrum presents now a non-vanishing absorption peak at $\omega\approx-\omega_0$.
For voltage $V=4.5\omega_0$ and electromechanical coupling $\epsilon_P=5.0\omega_0$, we find the computed ratio $S_{em}(\omega)/S_{abs}(\omega)\approx 2.0$, which
is consistent with having heating of the mechanical oscillator, with an average number of phonons $\bar{n}\approx 1.0$ and an effective
temperature $T_{eff}\approx 1.5 \omega_0$ [see Fig.~\ref{fig:P_x_Voltage}-(a)].
\\
Moreover, we find an overall good agreement between the Born-Markov and secular approximation results.
Some differences emerge in the tails of the three main peaks of the spectrum.
A zoom onto the symmetrized spectrum close to the emission peak at $\omega\approx\omega_0$ is plotted on Fig.~\ref{fig:S_xx_Full_1}-(b).
It is shown there that the Lamb-shift terms generated by Eq.\refe{Correlation_Function} 
are responsible for a weak softening of the mechanical mode frequency that is otherwise neglected within the secular approximation.
\\

Finally, we investigate on Fig.~\ref{fig:FWHM_x_Voltage} the dependence of the FWHM $\Delta \omega_x$ for the displacement fluctuation spectrum with both
bias-voltage and electromechanical coupling.
Upon increasing the bias-voltage from $V=0.2\omega_0$ to $V=4.5\omega_0$, we show that the maximum of the FWHM $\Delta\omega^{max}_x$
is shifted toward higher values of $\epsilon_P$.
We attribute this effect to the entering of additional vibronic sidebands into the bias-voltage window, which
opens new electric channels for decoherence and dephasing $\Delta\omega^{max}_\phi$ of the mechanical oscillator.
The distribution $\Delta \omega_x$ as a function of $\epsilon_P$ becomes also much broader at higher voltages compared to the low-bias case.
This implies more sensitivity of the mechanical oscillator to decoherence.
Indeed the unavoidable fluctuations in experimental $\epsilon_P$ values due to disorder, will induce 
an enhanced inhomogeneous broadening of the spectral line through the flat dependence of $\Delta \omega_x$ with $\epsilon_P$.
%

\subsection{Phase diagram}
%
%
%
\begin{figure}[tbh]
  \includegraphics[width=1.0\linewidth]{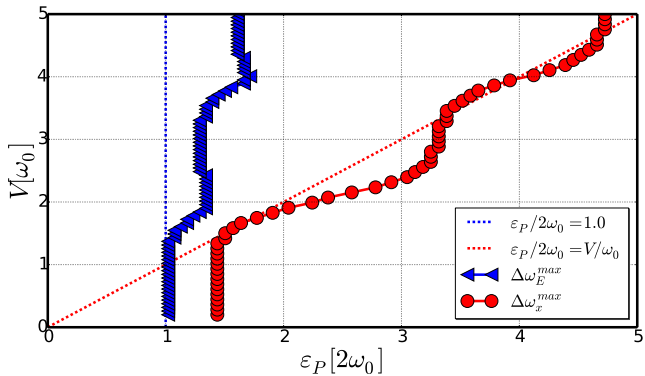}
  \caption{
    Locus of the points $(\epsilon_P/2,V)$ of maxima in the FHWM $\Delta \omega^{max}_E$ (blue triangles) and $\Delta \omega^{max}_x$ (red circles).
    Numerical results using the secular approximation developed in Sec.~\ref{Secular_Approximation}.
    Red dashed curve: critical coupling for the current blockade transition in the classical regime $\epsilon_P/2=V$.
    Blue dashed curve: critical coupling for the current blockade transition induced by ground-state quantum fluctuations $\epsilon_P=2.0\omega_0$.
    Chosen parameters: $\Gamma=0.05\omega_0$, $\tilde{\varepsilon}_0=0$ and $T = 0.1 \omega_0$.
  }
\label{fig:Phase_Diagram_xx}
\end{figure}
%
%
We summarize our findings in a phase diagram represented in Fig.~\ref{fig:Phase_Diagram_xx}.
The locus of the points $(\epsilon_P/2,V)$ of maxima in the FHWM $\Delta \omega^{max}_E$ is plotted with blue triangles.
For $0<V/2<\omega_0$, namely when the mechanical oscillator is close to its quantum ground state,
we find that the position of those maxima is independent of voltage and located at values
of the electromechanical coupling $g^2= \epsilon_P/2\omega_0=1$ (blue-dashed curve).
This is consistent with the results of Sec.~\ref{Dissipation_of_energy}, for which
the point of maximum energy dissipation coincides with the transition from a monostable mechanical oscillator (for $g^2<1$)
to a bistable one (for $g^2>1$). 
Upon increasing voltage above the first vibrational sideband ($\omega_0<V/2<2\omega_0$), the location of the maxima
increases toward a larger voltage-independent value $\epsilon_P/2\omega_0 \approx 1.3$.
Consistently with Sec.~\ref{Heating_mechanical_oscillator}, we assign this increased energy dissipation rate to the opening
of new inelastic electronic channels, each time a vibron sideband ($n$) is excited by the bias voltage ($V/2 > n\omega_0$).
Finally, the corresponding curve representing the location of the maxima in the FWHM $\Delta \omega^{max}_x$ 
is presented with red circles.
The obtained red curve is always on the right of the previous blue curve.
This is consistent with the analysis performed in Sec.~\ref{Oscillator_decoherence_time}, for which it is shown that
the decoherence rate of the mechanical oscillator is larger than the dissipation rate of energy because of
additional dephasing induced by elastically tunneling electrons.
A low voltages ($0<V/2<\omega_0$) the red curve is voltage independent and pinned at electromechanical coupling $g^2=\epsilon_P/2\omega_0=\sqrt{2}>1$.
This coincides with the value of $g^2$ maximizing the decoherence rate.
Upon increasing voltage to the range $\omega_0<V/2<2\omega_0$, we find that the locus of maximum decoherence increases in a step-like manner
towards a larger value of the coupling strength $\epsilon_P/2\omega_0\approx 3.3$.
This corresponds to the entering of a new vibron sideband $n=1$, which increases both the dissipation rate (through inelastic transitions) and the
dephasing rate (through enhanced elastic transitions).
Interestingly, we find that upon increasing sufficiently the bias voltage, the location of the maxima in the FWHM $\Delta \omega^{max}_x$ gets closer
to the red-dashed curve $V=\epsilon_P/2$.
We give a simple explanation of this phenomenon based on a semi-classical argument (at high-voltage indeed, many phonons populate the mechanical oscillator,
which becomes semi-classical).
The argument follows closely the analysis of the current-blockade phenomena in semi-classical mechanical oscillators \cite{galperin_hysteresis_2005,mozyrsky_intermittent_2006,pistolesi_current_2007,pistolesi_self-consistent_2008}.
We use for this the Hamiltonian written in Eq.\refe{Hamiltonian2}.
The tunneling electrons on the dot induce a back-action force on the mechanical oscillator
$\left\langle F \right\rangle = - F_0 \left\langle n_d \right\rangle$.
This back-action force in turn produces a shift of the oscillator equilibrium 
position $\Delta X_{eq}=- F_0/k \left\langle n_d \right\rangle$.
The work performed by the force $\left\langle F \right\rangle$ for displacing the equilibrium position of the oscillator by an amount $\Delta X_{eq}$
can be interpreted as a reorganization energy of the dot-level position
$\Delta \epsilon_0 = -\left\langle F \right\rangle\Delta X_{eq}$.
At half-filling ($\left\langle n_d \right\rangle=1/2$), we obtain $\Delta \epsilon_0 = -\epsilon_P/4$.
If $\Delta \epsilon_0$ is smaller than $-V/2$, namely that $\epsilon_P/2>V$,
the dot-level position is effectively shifted away from the conduction window and the current is blocked.
The critical value for this transition happens at $\epsilon_P/2=V$ (red-dashed curve) 
and coincides at high-voltage with the
transition from a monostable to a bistable state of the semi-classical oscillator.
%

\section{Conclusion}

It is well known that  a nano-electromechanical oscillator in the regime $\Gamma \ll T \ll \omega_0$
for large coupling constant $g^2=\epsilon_P/2\omega_0$ enters in the so-called Franck-Condon blockade regime. 
We have shown that the blockade sets in with a behavior similar to what observed in the semi-classical case,
namely the appearance of a double maximum in the probability distribution for the position of the oscillator. 
This property can be interpreted as a mechanical bistability present also in the quantum regime, even if one cannot define an effective potential as in the classical case.
At $T\ll \omega_0$ the transition point can be identified for $\epsilon_P=2\omega_0$ ($g^2=1$)  [see Fig.~\ref{fig:P_x}] 
while in presence of bias voltage $\epsilon_P/2 \approx V $ [see Fig.~\ref{fig:P_x_Voltage}].  
This is similar to what is found in the classical case for $\Gamma\gg T \gg \omega_0$ 
for which the transition happens at 
$\epsilon_P=\pi \Gamma$ \cite{PhysRevLett.115.206802,PhysRevB.94.125417}, 
with a smoothing given by thermal or non-equilibrium fluctuations.
Despite the similarity, the main difference between the two regimes is that in the classical case the transition is 
controlled by the change of the effective potential, while in the quantum case the quantum fluctuations are responsible for the disappearance of the bistability.
In analogy with the classical case we have investigated the displacement and energy fluctuation spectra. 
In the case of a quantum and fast oscillator, the lineshape of the spectra remains Lorentzian.
Somewhat surprisingly we find that the width $\Delta \omega$ of both is not monotonic and that the spectra are maximal
exactly at the bistable transition for $\Delta \omega_E$ and at slightly stronger coupling ($\epsilon_P=2\sqrt{2}\omega_0$) for  
$\Delta \omega_x$.
We presented a simple analytical analysis valid at low excitation probability of the oscillator (low $T$ or $V$) 
that allows to understand the origin of these widths.
In the weak coupling limit this is simply the lowest non-vanishing order in the perturbative expansion which shows a 
quadratic behavior. 
In the strong coupling limit the suppression of the tunneling due to the Franck-Condon terms suppresses also dissipation and decoherence, that can only be mediated by the electrons. 
Like in the classical case the width of the displacement spectrum (decoherence rate)
is larger than (half) $\Delta \omega_E$, the typical dissipation rate. 
In the quantum case the origin is not the 
non-linear effective potential, but the elastic transitions, that introduce decoherence without dissipation. 
We also investigate the same quantities as a function of the bias voltage and found that the 
dissipation and decoherence rates increase abruptly, each time a new vibrational side-band
enters into the conduction window, namely when $V/2$ becomes larger than a multiple 
of the mechanical frequency $\omega_0$. 
This gives rise to a phase diagram recovering the semi-classical limit for the current-blockade transition
(occurring when $\epsilon_P \gg V$)\cite{pistolesi_self-consistent_2008} at sufficiently high-voltages ($V\gg\omega_0$).
We found that the Wigner distribution of such an oscillator 
even close to its quantum ground state or to the threshold for inelastic transitions does not exhibit negative values.
This is due to the incoherent nature of the electron tunneling in this regime.

In conclusion we have found that the classical picture applies, at least partially, also in the quantum regime.
This scenario could be observed in high frequency mechanical oscillators. 
From the theoretical point of view other questions are still open.
It would be interesting to extend the present work to regimes of higher tunneling rates $\Gamma/T$, taking into account corrections induced by cotunneling of electrons.
Addressing the fate of the bistability transition in the regime of both coherent tunneling of electrons and quantum mechanical oscillator is still an opened theoretical issue even if recently a mapping has been established to an effective Kondo problem in the limit of slow oscillator in equilibrium \cite{PhysRevB.91.155306}.
Finally, it would be of interest for future works to investigate the possibility of generating non-classical states of the
mechanical oscillator by parametric driving \cite{PhysRevLett.67.699},
or by a suitable coupling of the nanotube mechanical oscillator to superconducting electrodes \cite{schneider2012coupling,PhysRevLett.97.196801,PhysRevB.86.155448}.
These results and perspectives contribute to show that non-trivial physical behavior arises
from the strong coupling between tunneling electrons and a well-controlled mechanical 
degree of freedom of the oscillator. 

\begin{acknowledgments}
We thank G. Micchi for comments on the manuscript. 
We acknowledge support of Conseil Regional de la Nouvelle Aquitaine. 
\end{acknowledgments}

\bibliography{Quantum_Oscillator}

\begin{thebibliography}{52}%
\makeatletter
\providecommand \@ifxundefined [1]{%
 \@ifx{#1\undefined}
}%
\providecommand \@ifnum [1]{%
 \ifnum #1\expandafter \@firstoftwo
 \else \expandafter \@secondoftwo
 \fi
}%
\providecommand \@ifx [1]{%
 \ifx #1\expandafter \@firstoftwo
 \else \expandafter \@secondoftwo
 \fi
}%
\providecommand \natexlab [1]{#1}%
\providecommand \enquote  [1]{``#1''}%
\providecommand \bibnamefont  [1]{#1}%
\providecommand \bibfnamefont [1]{#1}%
\providecommand \citenamefont [1]{#1}%
\providecommand \href@noop [0]{\@secondoftwo}%
\providecommand \href [0]{\begingroup \@sanitize@url \@href}%
\providecommand \@href[1]{\@@startlink{#1}\@@href}%
\providecommand \@@href[1]{\endgroup#1\@@endlink}%
\providecommand \@sanitize@url [0]{\catcode `\\12\catcode `\$12\catcode
  `\&12\catcode `\#12\catcode `\^12\catcode `\_12\catcode `\%12\relax}%
\providecommand \@@startlink[1]{}%
\providecommand \@@endlink[0]{}%
\providecommand \url  [0]{\begingroup\@sanitize@url \@url }%
\providecommand \@url [1]{\endgroup\@href {#1}{\urlprefix }}%
\providecommand \urlprefix  [0]{URL }%
\providecommand \Eprint [0]{\href }%
\providecommand \doibase [0]{http://dx.doi.org/}%
\providecommand \selectlanguage [0]{\@gobble}%
\providecommand \bibinfo  [0]{\@secondoftwo}%
\providecommand \bibfield  [0]{\@secondoftwo}%
\providecommand \translation [1]{[#1]}%
\providecommand \BibitemOpen [0]{}%
\providecommand \bibitemStop [0]{}%
\providecommand \bibitemNoStop [0]{.\EOS\space}%
\providecommand \EOS [0]{\spacefactor3000\relax}%
\providecommand \BibitemShut  [1]{\csname bibitem#1\endcsname}%
\let\auto@bib@innerbib\@empty
\bibitem [{\citenamefont {Imboden}\ and\ \citenamefont
  {Mohanty}(2014)}]{IMBODEN201489}%
  \BibitemOpen
  \bibfield  {author} {\bibinfo {author} {\bibfnamefont {M.}~\bibnamefont
  {Imboden}}\ and\ \bibinfo {author} {\bibfnamefont {P.}~\bibnamefont
  {Mohanty}},\ }\href {\doibase https://doi.org/10.1016/j.physrep.2013.09.003}
  {\bibfield  {journal} {\bibinfo  {journal} {Physics Reports}\ }\textbf
  {\bibinfo {volume} {534}},\ \bibinfo {pages} {89} (\bibinfo {year}
  {2014})}\BibitemShut {NoStop}%
\bibitem [{\citenamefont {Sazonova}\ \emph {et~al.}(2004)\citenamefont
  {Sazonova}, \citenamefont {Yaish}, \citenamefont {\"Ust\"unel}, \citenamefont
  {Roundy}, \citenamefont {Arias},\ and\ \citenamefont
  {McEuen}}]{sazonova_tunable_2004}%
  \BibitemOpen
  \bibfield  {author} {\bibinfo {author} {\bibfnamefont {V.}~\bibnamefont
  {Sazonova}}, \bibinfo {author} {\bibfnamefont {Y.}~\bibnamefont {Yaish}},
  \bibinfo {author} {\bibfnamefont {H.}~\bibnamefont {\"Ust\"unel}}, \bibinfo
  {author} {\bibfnamefont {D.}~\bibnamefont {Roundy}}, \bibinfo {author}
  {\bibfnamefont {T.~A.}\ \bibnamefont {Arias}}, \ and\ \bibinfo {author}
  {\bibfnamefont {P.~L.}\ \bibnamefont {McEuen}},\ }\href
  {http://www.nature.com/nature/journal/v431/n7006/abs/nature02905.html}
  {\bibfield  {journal} {\bibinfo  {journal} {Nature}\ }\textbf {\bibinfo
  {volume} {431}},\ \bibinfo {pages} {284} (\bibinfo {year}
  {2004})}\BibitemShut {NoStop}%
\bibitem [{\citenamefont {Lassagne}\ \emph {et~al.}(2009)\citenamefont
  {Lassagne}, \citenamefont {Tarakanov}, \citenamefont {Kinaret}, \citenamefont
  {Garcia-Sanchez},\ and\ \citenamefont {Bachtold}}]{lassagne_coupling_2009}%
  \BibitemOpen
  \bibfield  {author} {\bibinfo {author} {\bibfnamefont {B.}~\bibnamefont
  {Lassagne}}, \bibinfo {author} {\bibfnamefont {Y.}~\bibnamefont {Tarakanov}},
  \bibinfo {author} {\bibfnamefont {J.}~\bibnamefont {Kinaret}}, \bibinfo
  {author} {\bibfnamefont {D.}~\bibnamefont {Garcia-Sanchez}}, \ and\ \bibinfo
  {author} {\bibfnamefont {A.}~\bibnamefont {Bachtold}},\ }\href {\doibase
  10.1126/science.1174290} {\bibfield  {journal} {\bibinfo  {journal}
  {Science}\ }\textbf {\bibinfo {volume} {325}},\ \bibinfo {pages} {1107}
  (\bibinfo {year} {2009})}\BibitemShut {NoStop}%
\bibitem [{\citenamefont {Steele}\ \emph {et~al.}(2009)\citenamefont {Steele},
  \citenamefont {H\"uttel}, \citenamefont {Witkamp}, \citenamefont {Poot},
  \citenamefont {Meerwaldt}, \citenamefont {Kouwenhoven},\ and\ \citenamefont
  {van~der Zant}}]{steele_strong_2009}%
  \BibitemOpen
  \bibfield  {author} {\bibinfo {author} {\bibfnamefont {G.~A.}\ \bibnamefont
  {Steele}}, \bibinfo {author} {\bibfnamefont {A.~K.}\ \bibnamefont
  {H\"uttel}}, \bibinfo {author} {\bibfnamefont {B.}~\bibnamefont {Witkamp}},
  \bibinfo {author} {\bibfnamefont {M.}~\bibnamefont {Poot}}, \bibinfo {author}
  {\bibfnamefont {H.~B.}\ \bibnamefont {Meerwaldt}}, \bibinfo {author}
  {\bibfnamefont {L.~P.}\ \bibnamefont {Kouwenhoven}}, \ and\ \bibinfo {author}
  {\bibfnamefont {H.~S.}\ \bibnamefont {van~der Zant}},\ }\href
  {http://www.sciencemag.org/content/325/5944/1103.short} {\bibfield  {journal}
  {\bibinfo  {journal} {Science}\ }\textbf {\bibinfo {volume} {325}},\ \bibinfo
  {pages} {1103} (\bibinfo {year} {2009})}\BibitemShut {NoStop}%
\bibitem [{\citenamefont {Gouttenoire}\ \emph {et~al.}(2010)\citenamefont
  {Gouttenoire}, \citenamefont {Barois}, \citenamefont {Perisanu},
  \citenamefont {Leclercq}, \citenamefont {Purcell}, \citenamefont {Vincent},\
  and\ \citenamefont {Ayari}}]{SMLL:SMLL200901984}%
  \BibitemOpen
  \bibfield  {author} {\bibinfo {author} {\bibfnamefont {V.}~\bibnamefont
  {Gouttenoire}}, \bibinfo {author} {\bibfnamefont {T.}~\bibnamefont {Barois}},
  \bibinfo {author} {\bibfnamefont {S.}~\bibnamefont {Perisanu}}, \bibinfo
  {author} {\bibfnamefont {J.-L.}\ \bibnamefont {Leclercq}}, \bibinfo {author}
  {\bibfnamefont {S.~T.}\ \bibnamefont {Purcell}}, \bibinfo {author}
  {\bibfnamefont {P.}~\bibnamefont {Vincent}}, \ and\ \bibinfo {author}
  {\bibfnamefont {A.}~\bibnamefont {Ayari}},\ }\href {\doibase
  10.1002/smll.200901984} {\bibfield  {journal} {\bibinfo  {journal} {Small}\
  }\textbf {\bibinfo {volume} {6}},\ \bibinfo {pages} {1060} (\bibinfo {year}
  {2010})}\BibitemShut {NoStop}%
\bibitem [{\citenamefont {Laird}\ \emph {et~al.}(2011)\citenamefont {Laird},
  \citenamefont {Pei}, \citenamefont {Tang}, \citenamefont {Steele},\ and\
  \citenamefont {Kouwenhoven}}]{laird_high_2011}%
  \BibitemOpen
  \bibfield  {author} {\bibinfo {author} {\bibfnamefont {E.~A.}\ \bibnamefont
  {Laird}}, \bibinfo {author} {\bibfnamefont {F.}~\bibnamefont {Pei}}, \bibinfo
  {author} {\bibfnamefont {W.}~\bibnamefont {Tang}}, \bibinfo {author}
  {\bibfnamefont {G.~A.}\ \bibnamefont {Steele}}, \ and\ \bibinfo {author}
  {\bibfnamefont {L.~P.}\ \bibnamefont {Kouwenhoven}},\ }\href
  {http://pubs.acs.org/doi/abs/10.1021/nl203279v} {\bibfield  {journal}
  {\bibinfo  {journal} {Nano letters}\ }\textbf {\bibinfo {volume} {12}},\
  \bibinfo {pages} {193} (\bibinfo {year} {2011})}\BibitemShut {NoStop}%
\bibitem [{\citenamefont {Chaste}\ \emph {et~al.}(2011)\citenamefont {Chaste},
  \citenamefont {Sledzinska}, \citenamefont {Zdrojek}, \citenamefont {Moser},\
  and\ \citenamefont
  {Bachtold}}]{High_frequency_nanotube_mechanical_resonators}%
  \BibitemOpen
  \bibfield  {author} {\bibinfo {author} {\bibfnamefont {J.}~\bibnamefont
  {Chaste}}, \bibinfo {author} {\bibfnamefont {M.}~\bibnamefont {Sledzinska}},
  \bibinfo {author} {\bibfnamefont {M.}~\bibnamefont {Zdrojek}}, \bibinfo
  {author} {\bibfnamefont {J.}~\bibnamefont {Moser}}, \ and\ \bibinfo {author}
  {\bibfnamefont {A.}~\bibnamefont {Bachtold}},\ }\href {\doibase
  10.1063/1.3663630} {\bibfield  {journal} {\bibinfo  {journal} {Applied
  Physics Letters}\ }\textbf {\bibinfo {volume} {99}},\ \bibinfo {pages}
  {213502} (\bibinfo {year} {2011})}\BibitemShut {NoStop}%
\bibitem [{\citenamefont {Chaste}\ \emph {et~al.}(2012)\citenamefont {Chaste},
  \citenamefont {Eichler}, \citenamefont {Moser}, \citenamefont {Ceballos},
  \citenamefont {Rurali},\ and\ \citenamefont
  {Bachtold}}]{chaste_nanomechanical_2012}%
  \BibitemOpen
  \bibfield  {author} {\bibinfo {author} {\bibfnamefont {J.}~\bibnamefont
  {Chaste}}, \bibinfo {author} {\bibfnamefont {A.}~\bibnamefont {Eichler}},
  \bibinfo {author} {\bibfnamefont {J.}~\bibnamefont {Moser}}, \bibinfo
  {author} {\bibfnamefont {G.}~\bibnamefont {Ceballos}}, \bibinfo {author}
  {\bibfnamefont {R.}~\bibnamefont {Rurali}}, \ and\ \bibinfo {author}
  {\bibfnamefont {A.}~\bibnamefont {Bachtold}},\ }\href {\doibase
  10.1038/nnano.2012.42} {\bibfield  {journal} {\bibinfo  {journal} {Nature
  Nanotechnology}\ }\textbf {\bibinfo {volume} {7}},\ \bibinfo {pages} {301}
  (\bibinfo {year} {2012})}\BibitemShut {NoStop}%
\bibitem [{\citenamefont {Ganzhorn}\ and\ \citenamefont
  {Wernsdorfer}(2012)}]{ganzhorn_dynamics_2012}%
  \BibitemOpen
  \bibfield  {author} {\bibinfo {author} {\bibfnamefont {M.}~\bibnamefont
  {Ganzhorn}}\ and\ \bibinfo {author} {\bibfnamefont {W.}~\bibnamefont
  {Wernsdorfer}},\ }\href {\doibase 10.1103/PhysRevLett.108.175502} {\bibfield
  {journal} {\bibinfo  {journal} {Phys. Rev. Lett.}\ }\textbf {\bibinfo
  {volume} {108}},\ \bibinfo {pages} {175502} (\bibinfo {year}
  {2012})}\BibitemShut {NoStop}%
\bibitem [{\citenamefont {Meerwaldt}\ \emph {et~al.}(2012)\citenamefont
  {Meerwaldt}, \citenamefont {Labadze}, \citenamefont {Schneider},
  \citenamefont {Taspinar}, \citenamefont {Blanter}, \citenamefont {van~der
  Zant},\ and\ \citenamefont {Steele}}]{PhysRevB.86.115454}%
  \BibitemOpen
  \bibfield  {author} {\bibinfo {author} {\bibfnamefont {H.~B.}\ \bibnamefont
  {Meerwaldt}}, \bibinfo {author} {\bibfnamefont {G.}~\bibnamefont {Labadze}},
  \bibinfo {author} {\bibfnamefont {B.~H.}\ \bibnamefont {Schneider}}, \bibinfo
  {author} {\bibfnamefont {A.}~\bibnamefont {Taspinar}}, \bibinfo {author}
  {\bibfnamefont {Y.~M.}\ \bibnamefont {Blanter}}, \bibinfo {author}
  {\bibfnamefont {H.~S.~J.}\ \bibnamefont {van~der Zant}}, \ and\ \bibinfo
  {author} {\bibfnamefont {G.~A.}\ \bibnamefont {Steele}},\ }\href {\doibase
  10.1103/PhysRevB.86.115454} {\bibfield  {journal} {\bibinfo  {journal} {Phys.
  Rev. B}\ }\textbf {\bibinfo {volume} {86}},\ \bibinfo {pages} {115454}
  (\bibinfo {year} {2012})}\BibitemShut {NoStop}%
\bibitem [{\citenamefont {Wang}\ and\ \citenamefont
  {Pistolesi}(2017)}]{PhysRevB.95.035410}%
  \BibitemOpen
  \bibfield  {author} {\bibinfo {author} {\bibfnamefont {Y.}~\bibnamefont
  {Wang}}\ and\ \bibinfo {author} {\bibfnamefont {F.}~\bibnamefont
  {Pistolesi}},\ }\href {\doibase 10.1103/PhysRevB.95.035410} {\bibfield
  {journal} {\bibinfo  {journal} {Phys. Rev. B}\ }\textbf {\bibinfo {volume}
  {95}},\ \bibinfo {pages} {035410} (\bibinfo {year} {2017})}\BibitemShut
  {NoStop}%
\bibitem [{\citenamefont {Waissman}\ \emph {et~al.}(2013)\citenamefont
  {Waissman}, \citenamefont {Honig}, \citenamefont {Pecker}, \citenamefont
  {Benyamini}, \citenamefont {Hamo},\ and\ \citenamefont
  {Ilani}}]{waissman2013realization}%
  \BibitemOpen
  \bibfield  {author} {\bibinfo {author} {\bibfnamefont {J.}~\bibnamefont
  {Waissman}}, \bibinfo {author} {\bibfnamefont {M.}~\bibnamefont {Honig}},
  \bibinfo {author} {\bibfnamefont {S.}~\bibnamefont {Pecker}}, \bibinfo
  {author} {\bibfnamefont {A.}~\bibnamefont {Benyamini}}, \bibinfo {author}
  {\bibfnamefont {A.}~\bibnamefont {Hamo}}, \ and\ \bibinfo {author}
  {\bibfnamefont {S.}~\bibnamefont {Ilani}},\ }\href@noop {} {\bibfield
  {journal} {\bibinfo  {journal} {Nature nanotechnology}\ }\textbf {\bibinfo
  {volume} {8}},\ \bibinfo {pages} {569} (\bibinfo {year} {2013})}\BibitemShut
  {NoStop}%
\bibitem [{\citenamefont {Benyamini}\ \emph {et~al.}(2014)\citenamefont
  {Benyamini}, \citenamefont {Hamo}, \citenamefont {Kusminskiy}, \citenamefont
  {von Oppen},\ and\ \citenamefont {Ilani}}]{benyamini_real-space_2014}%
  \BibitemOpen
  \bibfield  {author} {\bibinfo {author} {\bibfnamefont {A.}~\bibnamefont
  {Benyamini}}, \bibinfo {author} {\bibfnamefont {A.}~\bibnamefont {Hamo}},
  \bibinfo {author} {\bibfnamefont {S.~V.}\ \bibnamefont {Kusminskiy}},
  \bibinfo {author} {\bibfnamefont {F.}~\bibnamefont {von Oppen}}, \ and\
  \bibinfo {author} {\bibfnamefont {S.}~\bibnamefont {Ilani}},\ }\href
  {\doibase 10.1038/nphys2842} {\bibfield  {journal} {\bibinfo  {journal}
  {Nature Physics}\ }\textbf {\bibinfo {volume} {10}},\ \bibinfo {pages} {151}
  (\bibinfo {year} {2014})}\BibitemShut {NoStop}%
\bibitem [{\citenamefont {Micchi}\ \emph {et~al.}(2015)\citenamefont {Micchi},
  \citenamefont {Avriller},\ and\ \citenamefont
  {Pistolesi}}]{PhysRevLett.115.206802}%
  \BibitemOpen
  \bibfield  {author} {\bibinfo {author} {\bibfnamefont {G.}~\bibnamefont
  {Micchi}}, \bibinfo {author} {\bibfnamefont {R.}~\bibnamefont {Avriller}}, \
  and\ \bibinfo {author} {\bibfnamefont {F.}~\bibnamefont {Pistolesi}},\ }\href
  {\doibase 10.1103/PhysRevLett.115.206802} {\bibfield  {journal} {\bibinfo
  {journal} {Phys. Rev. Lett.}\ }\textbf {\bibinfo {volume} {115}},\ \bibinfo
  {pages} {206802} (\bibinfo {year} {2015})}\BibitemShut {NoStop}%
\bibitem [{\citenamefont {Moser}\ \emph {et~al.}(2013)\citenamefont {Moser},
  \citenamefont {G{\"u}ttinger}, \citenamefont {Eichler}, \citenamefont
  {Esplandiu}, \citenamefont {Liu}, \citenamefont {Dykman},\ and\ \citenamefont
  {Bachtold}}]{moser2013ultrasensitive}%
  \BibitemOpen
  \bibfield  {author} {\bibinfo {author} {\bibfnamefont {J.}~\bibnamefont
  {Moser}}, \bibinfo {author} {\bibfnamefont {J.}~\bibnamefont
  {G{\"u}ttinger}}, \bibinfo {author} {\bibfnamefont {A.}~\bibnamefont
  {Eichler}}, \bibinfo {author} {\bibfnamefont {M.~J.}\ \bibnamefont
  {Esplandiu}}, \bibinfo {author} {\bibfnamefont {D.}~\bibnamefont {Liu}},
  \bibinfo {author} {\bibfnamefont {M.}~\bibnamefont {Dykman}}, \ and\ \bibinfo
  {author} {\bibfnamefont {A.}~\bibnamefont {Bachtold}},\ }\href {\doibase
  10.1038/nnano.2013.97} {\bibfield  {journal} {\bibinfo  {journal} {Nature
  nanotechnology}\ }\textbf {\bibinfo {volume} {8}},\ \bibinfo {pages} {493}
  (\bibinfo {year} {2013})}\BibitemShut {NoStop}%
\bibitem [{\citenamefont {Moser}\ \emph {et~al.}(2014)\citenamefont {Moser},
  \citenamefont {Eichler}, \citenamefont {G\"uttinger}, \citenamefont
  {Dykman},\ and\ \citenamefont {Bachtold}}]{moser_nanotube_2014}%
  \BibitemOpen
  \bibfield  {author} {\bibinfo {author} {\bibfnamefont {J.}~\bibnamefont
  {Moser}}, \bibinfo {author} {\bibfnamefont {A.}~\bibnamefont {Eichler}},
  \bibinfo {author} {\bibfnamefont {J.}~\bibnamefont {G\"uttinger}}, \bibinfo
  {author} {\bibfnamefont {M.~I.}\ \bibnamefont {Dykman}}, \ and\ \bibinfo
  {author} {\bibfnamefont {A.}~\bibnamefont {Bachtold}},\ }\href {\doibase
  10.1038/nnano.2014.234} {\bibfield  {journal} {\bibinfo  {journal} {Nat
  Nano}\ }\textbf {\bibinfo {volume} {9}},\ \bibinfo {pages} {1007} (\bibinfo
  {year} {2014})}\BibitemShut {NoStop}%
\bibitem [{\citenamefont {Micchi}\ \emph {et~al.}(2016)\citenamefont {Micchi},
  \citenamefont {Avriller},\ and\ \citenamefont
  {Pistolesi}}]{PhysRevB.94.125417}%
  \BibitemOpen
  \bibfield  {author} {\bibinfo {author} {\bibfnamefont {G.}~\bibnamefont
  {Micchi}}, \bibinfo {author} {\bibfnamefont {R.}~\bibnamefont {Avriller}}, \
  and\ \bibinfo {author} {\bibfnamefont {F.}~\bibnamefont {Pistolesi}},\ }\href
  {\doibase 10.1103/PhysRevB.94.125417} {\bibfield  {journal} {\bibinfo
  {journal} {Phys. Rev. B}\ }\textbf {\bibinfo {volume} {94}},\ \bibinfo
  {pages} {125417} (\bibinfo {year} {2016})}\BibitemShut {NoStop}%
\bibitem [{\citenamefont {Galperin}\ \emph {et~al.}(2005)\citenamefont
  {Galperin}, \citenamefont {Ratner},\ and\ \citenamefont
  {Nitzan}}]{galperin_hysteresis_2005}%
  \BibitemOpen
  \bibfield  {author} {\bibinfo {author} {\bibfnamefont {M.}~\bibnamefont
  {Galperin}}, \bibinfo {author} {\bibfnamefont {M.~A.}\ \bibnamefont
  {Ratner}}, \ and\ \bibinfo {author} {\bibfnamefont {A.}~\bibnamefont
  {Nitzan}},\ }\href {\doibase 10.1021/nl048216c} {\bibfield  {journal}
  {\bibinfo  {journal} {Nano Lett.}\ }\textbf {\bibinfo {volume} {5}},\
  \bibinfo {pages} {125} (\bibinfo {year} {2005})}\BibitemShut {NoStop}%
\bibitem [{\citenamefont {Mozyrsky}\ \emph {et~al.}(2006)\citenamefont
  {Mozyrsky}, \citenamefont {Hastings},\ and\ \citenamefont
  {Martin}}]{mozyrsky_intermittent_2006}%
  \BibitemOpen
  \bibfield  {author} {\bibinfo {author} {\bibfnamefont {D.}~\bibnamefont
  {Mozyrsky}}, \bibinfo {author} {\bibfnamefont {M.~B.}\ \bibnamefont
  {Hastings}}, \ and\ \bibinfo {author} {\bibfnamefont {I.}~\bibnamefont
  {Martin}},\ }\href {\doibase 10.1103/PhysRevB.73.035104} {\bibfield
  {journal} {\bibinfo  {journal} {Phys. Rev. B}\ }\textbf {\bibinfo {volume}
  {73}},\ \bibinfo {pages} {035104} (\bibinfo {year} {2006})}\BibitemShut
  {NoStop}%
\bibitem [{\citenamefont {Pistolesi}\ and\ \citenamefont
  {Labarthe}(2007)}]{pistolesi_current_2007}%
  \BibitemOpen
  \bibfield  {author} {\bibinfo {author} {\bibfnamefont {F.}~\bibnamefont
  {Pistolesi}}\ and\ \bibinfo {author} {\bibfnamefont {S.}~\bibnamefont
  {Labarthe}},\ }\href {\doibase 10.1103/PhysRevB.76.165317} {\bibfield
  {journal} {\bibinfo  {journal} {Phys. Rev. B}\ }\textbf {\bibinfo {volume}
  {76}},\ \bibinfo {pages} {165317} (\bibinfo {year} {2007})}\BibitemShut
  {NoStop}%
\bibitem [{\citenamefont {Pistolesi}\ \emph {et~al.}(2008)\citenamefont
  {Pistolesi}, \citenamefont {Blanter},\ and\ \citenamefont
  {Martin}}]{pistolesi_self-consistent_2008}%
  \BibitemOpen
  \bibfield  {author} {\bibinfo {author} {\bibfnamefont {F.}~\bibnamefont
  {Pistolesi}}, \bibinfo {author} {\bibfnamefont {Y.~M.}\ \bibnamefont
  {Blanter}}, \ and\ \bibinfo {author} {\bibfnamefont {I.}~\bibnamefont
  {Martin}},\ }\href {\doibase 10.1103/PhysRevB.78.085127} {\bibfield
  {journal} {\bibinfo  {journal} {Phys. Rev. B}\ }\textbf {\bibinfo {volume}
  {78}},\ \bibinfo {pages} {085127} (\bibinfo {year} {2008})}\BibitemShut
  {NoStop}%
\bibitem [{\citenamefont {Wang}\ \emph {et~al.}(2017)\citenamefont {Wang},
  \citenamefont {Micchi},\ and\ \citenamefont
  {Pistolesi}}]{0953-8984-29-46-465304}%
  \BibitemOpen
  \bibfield  {author} {\bibinfo {author} {\bibfnamefont {Y.}~\bibnamefont
  {Wang}}, \bibinfo {author} {\bibfnamefont {G.}~\bibnamefont {Micchi}}, \ and\
  \bibinfo {author} {\bibfnamefont {F.}~\bibnamefont {Pistolesi}},\ }\href@noop
  {} {\bibfield  {journal} {\bibinfo  {journal} {Journal of Physics: Condensed
  Matter}\ }\textbf {\bibinfo {volume} {29}},\ \bibinfo {pages} {465304}
  (\bibinfo {year} {2017})}\BibitemShut {NoStop}%
\bibitem [{\citenamefont {Koch}\ and\ \citenamefont {von
  Oppen}(2005)}]{koch_franck-condon_2005}%
  \BibitemOpen
  \bibfield  {author} {\bibinfo {author} {\bibfnamefont {J.}~\bibnamefont
  {Koch}}\ and\ \bibinfo {author} {\bibfnamefont {F.}~\bibnamefont {von
  Oppen}},\ }\href {\doibase 10.1103/PhysRevLett.94.206804} {\bibfield
  {journal} {\bibinfo  {journal} {Phys. Rev. Lett.}\ }\textbf {\bibinfo
  {volume} {94}},\ \bibinfo {pages} {206804} (\bibinfo {year}
  {2005})}\BibitemShut {NoStop}%
\bibitem [{\citenamefont {Koch}\ \emph {et~al.}(2006)\citenamefont {Koch},
  \citenamefont {von Oppen},\ and\ \citenamefont {Andreev}}]{koch_theory_2006}%
  \BibitemOpen
  \bibfield  {author} {\bibinfo {author} {\bibfnamefont {J.}~\bibnamefont
  {Koch}}, \bibinfo {author} {\bibfnamefont {F.}~\bibnamefont {von Oppen}}, \
  and\ \bibinfo {author} {\bibfnamefont {A.~V.}\ \bibnamefont {Andreev}},\
  }\href {\doibase 10.1103/PhysRevB.74.205438} {\bibfield  {journal} {\bibinfo
  {journal} {Phys. Rev. B}\ }\textbf {\bibinfo {volume} {74}},\ \bibinfo
  {pages} {205438} (\bibinfo {year} {2006})}\BibitemShut {NoStop}%
\bibitem [{\citenamefont {Braig}\ and\ \citenamefont
  {Flensberg}(2003)}]{braig_vibrational_2003}%
  \BibitemOpen
  \bibfield  {author} {\bibinfo {author} {\bibfnamefont {S.}~\bibnamefont
  {Braig}}\ and\ \bibinfo {author} {\bibfnamefont {K.}~\bibnamefont
  {Flensberg}},\ }\href@noop {} {\bibfield  {journal} {\bibinfo  {journal}
  {Phys. Rev. B}\ }\textbf {\bibinfo {volume} {68}},\ \bibinfo {pages} {205324}
  (\bibinfo {year} {2003})}\BibitemShut {NoStop}%
\bibitem [{\citenamefont {Mitra}\ \emph {et~al.}(2004)\citenamefont {Mitra},
  \citenamefont {Aleiner},\ and\ \citenamefont {Millis}}]{mitra_phonon_2004}%
  \BibitemOpen
  \bibfield  {author} {\bibinfo {author} {\bibfnamefont {A.}~\bibnamefont
  {Mitra}}, \bibinfo {author} {\bibfnamefont {I.}~\bibnamefont {Aleiner}}, \
  and\ \bibinfo {author} {\bibfnamefont {A.~J.}\ \bibnamefont {Millis}},\
  }\href {\doibase 10.1103/PhysRevB.69.245302} {\bibfield  {journal} {\bibinfo
  {journal} {Phys. Rev. B}\ }\textbf {\bibinfo {volume} {69}},\ \bibinfo
  {pages} {245302} (\bibinfo {year} {2004})}\BibitemShut {NoStop}%
\bibitem [{\citenamefont {Leturcq}\ \emph {et~al.}(2009)\citenamefont
  {Leturcq}, \citenamefont {Stampfer}, \citenamefont {Inderbitzin},
  \citenamefont {Durrer}, \citenamefont {Hierold}, \citenamefont {Mariani},
  \citenamefont {Schultz}, \citenamefont {von Oppen},\ and\ \citenamefont
  {Ensslin}}]{leturcq_franckcondon_2009}%
  \BibitemOpen
  \bibfield  {author} {\bibinfo {author} {\bibfnamefont {R.}~\bibnamefont
  {Leturcq}}, \bibinfo {author} {\bibfnamefont {C.}~\bibnamefont {Stampfer}},
  \bibinfo {author} {\bibfnamefont {K.}~\bibnamefont {Inderbitzin}}, \bibinfo
  {author} {\bibfnamefont {L.}~\bibnamefont {Durrer}}, \bibinfo {author}
  {\bibfnamefont {C.}~\bibnamefont {Hierold}}, \bibinfo {author} {\bibfnamefont
  {E.}~\bibnamefont {Mariani}}, \bibinfo {author} {\bibfnamefont {M.~G.}\
  \bibnamefont {Schultz}}, \bibinfo {author} {\bibfnamefont {F.}~\bibnamefont
  {von Oppen}}, \ and\ \bibinfo {author} {\bibfnamefont {K.}~\bibnamefont
  {Ensslin}},\ }\href {\doibase 10.1038/nphys1234} {\bibfield  {journal}
  {\bibinfo  {journal} {Nat Phys}\ }\textbf {\bibinfo {volume} {5}},\ \bibinfo
  {pages} {327} (\bibinfo {year} {2009})}\BibitemShut {NoStop}%
\bibitem [{\citenamefont {Burzur{\'\i}}\ \emph {et~al.}(2014)\citenamefont
  {Burzur{\'\i}}, \citenamefont {Yamamoto}, \citenamefont {Warnock},
  \citenamefont {Zhong}, \citenamefont {Park}, \citenamefont {Cornia},\ and\
  \citenamefont {van~der Zant}}]{burzuri2014franck}%
  \BibitemOpen
  \bibfield  {author} {\bibinfo {author} {\bibfnamefont {E.}~\bibnamefont
  {Burzur{\'\i}}}, \bibinfo {author} {\bibfnamefont {Y.}~\bibnamefont
  {Yamamoto}}, \bibinfo {author} {\bibfnamefont {M.}~\bibnamefont {Warnock}},
  \bibinfo {author} {\bibfnamefont {X.}~\bibnamefont {Zhong}}, \bibinfo
  {author} {\bibfnamefont {K.}~\bibnamefont {Park}}, \bibinfo {author}
  {\bibfnamefont {A.}~\bibnamefont {Cornia}}, \ and\ \bibinfo {author}
  {\bibfnamefont {H.~S.}\ \bibnamefont {van~der Zant}},\ }\href@noop {}
  {\bibfield  {journal} {\bibinfo  {journal} {Nano letters}\ }\textbf {\bibinfo
  {volume} {14}},\ \bibinfo {pages} {3191} (\bibinfo {year}
  {2014})}\BibitemShut {NoStop}%
\bibitem [{\citenamefont {H\"artle}\ and\ \citenamefont
  {Thoss}(2011)}]{PhysRevB.83.125419}%
  \BibitemOpen
  \bibfield  {author} {\bibinfo {author} {\bibfnamefont {R.}~\bibnamefont
  {H\"artle}}\ and\ \bibinfo {author} {\bibfnamefont {M.}~\bibnamefont
  {Thoss}},\ }\href {\doibase 10.1103/PhysRevB.83.125419} {\bibfield  {journal}
  {\bibinfo  {journal} {Phys. Rev. B}\ }\textbf {\bibinfo {volume} {83}},\
  \bibinfo {pages} {125419} (\bibinfo {year} {2011})}\BibitemShut {NoStop}%
\bibitem [{\citenamefont {Avriller}(2011)}]{0953-8984-23-10-105301}%
  \BibitemOpen
  \bibfield  {author} {\bibinfo {author} {\bibfnamefont {R.}~\bibnamefont
  {Avriller}},\ }\href {\doibase 10.1088/0953-8984/23/10/105301} {\bibfield
  {journal} {\bibinfo  {journal} {Journal of Physics: Condensed Matter}\
  }\textbf {\bibinfo {volume} {23}},\ \bibinfo {pages} {105301} (\bibinfo
  {year} {2011})}\BibitemShut {NoStop}%
\bibitem [{\citenamefont {Piovano}\ \emph {et~al.}(2011)\citenamefont
  {Piovano}, \citenamefont {Cavaliere}, \citenamefont {Paladino},\ and\
  \citenamefont {Sassetti}}]{PhysRevB.83.245311}%
  \BibitemOpen
  \bibfield  {author} {\bibinfo {author} {\bibfnamefont {G.}~\bibnamefont
  {Piovano}}, \bibinfo {author} {\bibfnamefont {F.}~\bibnamefont {Cavaliere}},
  \bibinfo {author} {\bibfnamefont {E.}~\bibnamefont {Paladino}}, \ and\
  \bibinfo {author} {\bibfnamefont {M.}~\bibnamefont {Sassetti}},\ }\href
  {\doibase 10.1103/PhysRevB.83.245311} {\bibfield  {journal} {\bibinfo
  {journal} {Phys. Rev. B}\ }\textbf {\bibinfo {volume} {83}},\ \bibinfo
  {pages} {245311} (\bibinfo {year} {2011})}\BibitemShut {NoStop}%
\bibitem [{\citenamefont {H\"ubener}\ and\ \citenamefont
  {Brandes}(2009)}]{PhysRevB.80.155437}%
  \BibitemOpen
  \bibfield  {author} {\bibinfo {author} {\bibfnamefont {H.}~\bibnamefont
  {H\"ubener}}\ and\ \bibinfo {author} {\bibfnamefont {T.}~\bibnamefont
  {Brandes}},\ }\href {\doibase 10.1103/PhysRevB.80.155437} {\bibfield
  {journal} {\bibinfo  {journal} {Phys. Rev. B}\ }\textbf {\bibinfo {volume}
  {80}},\ \bibinfo {pages} {155437} (\bibinfo {year} {2009})}\BibitemShut
  {NoStop}%
\bibitem [{\citenamefont {Lang}\ and\ \citenamefont
  {Firsov}(1963)}]{lang1963kinetic}%
  \BibitemOpen
  \bibfield  {author} {\bibinfo {author} {\bibfnamefont {I.}~\bibnamefont
  {Lang}}\ and\ \bibinfo {author} {\bibfnamefont {Y.~A.}\ \bibnamefont
  {Firsov}},\ }\href@noop {} {\bibfield  {journal} {\bibinfo  {journal} {Sov.
  Phys. JETP}\ }\textbf {\bibinfo {volume} {16}},\ \bibinfo {pages} {1301}
  (\bibinfo {year} {1963})}\BibitemShut {NoStop}%
\bibitem [{\citenamefont {Flensberg}(2003)}]{flensberg_tunneling_2003}%
  \BibitemOpen
  \bibfield  {author} {\bibinfo {author} {\bibfnamefont {K.}~\bibnamefont
  {Flensberg}},\ }\href {\doibase 10.1103/PhysRevB.68.205323} {\bibfield
  {journal} {\bibinfo  {journal} {Phys. Rev. B}\ }\textbf {\bibinfo {volume}
  {68}},\ \bibinfo {pages} {205323} (\bibinfo {year} {2003})}\BibitemShut
  {NoStop}%
\bibitem [{\citenamefont {Schlosshauer}(2007)}]{schlosshauer2007decoherence}%
  \BibitemOpen
  \bibfield  {author} {\bibinfo {author} {\bibfnamefont {M.~A.}\ \bibnamefont
  {Schlosshauer}},\ }\href@noop {} {\emph {\bibinfo {title} {Decoherence: and
  the quantum-to-classical transition}}}\ (\bibinfo  {publisher} {Springer
  Science \& Business Media},\ \bibinfo {year} {2007})\BibitemShut {NoStop}%
\bibitem [{\citenamefont {Rodrigues}\ and\ \citenamefont
  {Armour}(2005)}]{1367-2630-7-1-251}%
  \BibitemOpen
  \bibfield  {author} {\bibinfo {author} {\bibfnamefont {D.~A.}\ \bibnamefont
  {Rodrigues}}\ and\ \bibinfo {author} {\bibfnamefont {A.~D.}\ \bibnamefont
  {Armour}},\ }\href@noop {} {\bibfield  {journal} {\bibinfo  {journal} {New
  Journal of Physics}\ }\textbf {\bibinfo {volume} {7}},\ \bibinfo {pages}
  {251} (\bibinfo {year} {2005})}\BibitemShut {NoStop}%
\bibitem [{\citenamefont {Abramowitz}\ and\ \citenamefont
  {Stegun}(1964)}]{abramowitz1964handbook}%
  \BibitemOpen
  \bibfield  {author} {\bibinfo {author} {\bibfnamefont {M.}~\bibnamefont
  {Abramowitz}}\ and\ \bibinfo {author} {\bibfnamefont {I.~A.}\ \bibnamefont
  {Stegun}},\ }\href@noop {} {\emph {\bibinfo {title} {Handbook of mathematical
  functions: with formulas, graphs, and mathematical tables}}},\ \bibinfo
  {number} {55}\ (\bibinfo  {publisher} {Courier Corporation},\ \bibinfo {year}
  {1964})\BibitemShut {NoStop}%
\bibitem [{\citenamefont {Bevilacqua}\ \emph {et~al.}(2016)\citenamefont
  {Bevilacqua}, \citenamefont {Menichetti},\ and\ \citenamefont
  {Parravicini}}]{Bevilacqua2016}%
  \BibitemOpen
  \bibfield  {author} {\bibinfo {author} {\bibfnamefont {G.}~\bibnamefont
  {Bevilacqua}}, \bibinfo {author} {\bibfnamefont {G.}~\bibnamefont
  {Menichetti}}, \ and\ \bibinfo {author} {\bibfnamefont {G.~P.}\ \bibnamefont
  {Parravicini}},\ }\href {\doibase 10.1140/epjb/e2015-60730-0} {\bibfield
  {journal} {\bibinfo  {journal} {The European Physical Journal B}\ }\textbf
  {\bibinfo {volume} {89}},\ \bibinfo {pages} {3} (\bibinfo {year}
  {2016})}\BibitemShut {NoStop}%
\bibitem [{\citenamefont {Cohen-Tannoudji}\ \emph {et~al.}(1992)\citenamefont
  {Cohen-Tannoudji}, \citenamefont {Dupont-Roc}, \citenamefont {Grynberg},\
  and\ \citenamefont {Thickstun}}]{cohen1992atom}%
  \BibitemOpen
  \bibfield  {author} {\bibinfo {author} {\bibfnamefont {C.}~\bibnamefont
  {Cohen-Tannoudji}}, \bibinfo {author} {\bibfnamefont {J.}~\bibnamefont
  {Dupont-Roc}}, \bibinfo {author} {\bibfnamefont {G.}~\bibnamefont
  {Grynberg}}, \ and\ \bibinfo {author} {\bibfnamefont {P.}~\bibnamefont
  {Thickstun}},\ }\href@noop {} {\emph {\bibinfo {title} {Atom-photon
  interactions: basic processes and applications}}}\ (\bibinfo  {publisher}
  {Wiley Online Library},\ \bibinfo {year} {1992})\BibitemShut {NoStop}%
\bibitem [{\citenamefont {Kirton}\ \emph {et~al.}(2012)\citenamefont {Kirton},
  \citenamefont {Armour}, \citenamefont {Houzet},\ and\ \citenamefont
  {Pistolesi}}]{kirton_quantum_2012}%
  \BibitemOpen
  \bibfield  {author} {\bibinfo {author} {\bibfnamefont {P.~G.}\ \bibnamefont
  {Kirton}}, \bibinfo {author} {\bibfnamefont {A.~D.}\ \bibnamefont {Armour}},
  \bibinfo {author} {\bibfnamefont {M.}~\bibnamefont {Houzet}}, \ and\ \bibinfo
  {author} {\bibfnamefont {F.}~\bibnamefont {Pistolesi}},\ }\href {\doibase
  10.1103/PhysRevB.86.081305} {\bibfield  {journal} {\bibinfo  {journal} {Phys.
  Rev. B}\ }\textbf {\bibinfo {volume} {86}},\ \bibinfo {pages} {081305}
  (\bibinfo {year} {2012})}\BibitemShut {NoStop}%
\bibitem [{\citenamefont {Cohen-Tannoudji}\ \emph {et~al.}(1991)\citenamefont
  {Cohen-Tannoudji}, \citenamefont {Diu},\ and\ \citenamefont
  {Franck}}]{cohenquantummechanics}%
  \BibitemOpen
  \bibfield  {author} {\bibinfo {author} {\bibfnamefont {C.}~\bibnamefont
  {Cohen-Tannoudji}}, \bibinfo {author} {\bibfnamefont {B.}~\bibnamefont
  {Diu}}, \ and\ \bibinfo {author} {\bibfnamefont {L.}~\bibnamefont {Franck}},\
  }\href@noop {} {\emph {\bibinfo {title} {Quantum Mechanics: Volume One}}}\
  (\bibinfo  {publisher} {Wiley Online Library},\ \bibinfo {year}
  {1991})\BibitemShut {NoStop}%
\bibitem [{\citenamefont {Wigner}(1932)}]{PhysRev.40.749}%
  \BibitemOpen
  \bibfield  {author} {\bibinfo {author} {\bibfnamefont {E.}~\bibnamefont
  {Wigner}},\ }\href {\doibase 10.1103/PhysRev.40.749} {\bibfield  {journal}
  {\bibinfo  {journal} {Phys. Rev.}\ }\textbf {\bibinfo {volume} {40}},\
  \bibinfo {pages} {749} (\bibinfo {year} {1932})}\BibitemShut {NoStop}%
\bibitem [{\citenamefont {Lee}(1995)}]{LEE1995147}%
  \BibitemOpen
  \bibfield  {author} {\bibinfo {author} {\bibfnamefont {H.-W.}\ \bibnamefont
  {Lee}},\ }\href {\doibase 10.1016/0370-1573(95)00007-4} {\bibfield  {journal}
  {\bibinfo  {journal} {Physics Reports}\ }\textbf {\bibinfo {volume} {259}},\
  \bibinfo {pages} {147} (\bibinfo {year} {1995})}\BibitemShut {NoStop}%
\bibitem [{\citenamefont {Callen}\ and\ \citenamefont
  {Welton}(1951)}]{PhysRev.83.34}%
  \BibitemOpen
  \bibfield  {author} {\bibinfo {author} {\bibfnamefont {H.~B.}\ \bibnamefont
  {Callen}}\ and\ \bibinfo {author} {\bibfnamefont {T.~A.}\ \bibnamefont
  {Welton}},\ }\href {\doibase 10.1103/PhysRev.83.34} {\bibfield  {journal}
  {\bibinfo  {journal} {Phys. Rev.}\ }\textbf {\bibinfo {volume} {83}},\
  \bibinfo {pages} {34} (\bibinfo {year} {1951})}\BibitemShut {NoStop}%
\bibitem [{\citenamefont {Pistolesi}(2009)}]{pistolesi_cooling_2009}%
  \BibitemOpen
  \bibfield  {author} {\bibinfo {author} {\bibfnamefont {F.}~\bibnamefont
  {Pistolesi}},\ }\href {\doibase 10.1007/s10909-009-9867-1} {\bibfield
  {journal} {\bibinfo  {journal} {Journal of Low Temperature Physics}\ }\textbf
  {\bibinfo {volume} {154}},\ \bibinfo {pages} {199} (\bibinfo {year}
  {2009})}\BibitemShut {NoStop}%
\bibitem [{\citenamefont {Traverso~Ziani}\ \emph {et~al.}(2011)\citenamefont
  {Traverso~Ziani}, \citenamefont {Piovano}, \citenamefont {Cavaliere},\ and\
  \citenamefont {Sassetti}}]{traverso_ziani_electrical_2011}%
  \BibitemOpen
  \bibfield  {author} {\bibinfo {author} {\bibfnamefont {N.}~\bibnamefont
  {Traverso~Ziani}}, \bibinfo {author} {\bibfnamefont {G.}~\bibnamefont
  {Piovano}}, \bibinfo {author} {\bibfnamefont {F.}~\bibnamefont {Cavaliere}},
  \ and\ \bibinfo {author} {\bibfnamefont {M.}~\bibnamefont {Sassetti}},\
  }\href {\doibase 10.1103/PhysRevB.84.155423} {\bibfield  {journal} {\bibinfo
  {journal} {Phys. Rev. B}\ }\textbf {\bibinfo {volume} {84}},\ \bibinfo
  {pages} {155423} (\bibinfo {year} {2011})}\BibitemShut {NoStop}%
\bibitem [{\citenamefont {Armour}\ \emph {et~al.}(2004)\citenamefont {Armour},
  \citenamefont {Blencowe},\ and\ \citenamefont
  {Zhang}}]{armour_classical_2004}%
  \BibitemOpen
  \bibfield  {author} {\bibinfo {author} {\bibfnamefont {A.~D.}\ \bibnamefont
  {Armour}}, \bibinfo {author} {\bibfnamefont {M.~P.}\ \bibnamefont
  {Blencowe}}, \ and\ \bibinfo {author} {\bibfnamefont {Y.}~\bibnamefont
  {Zhang}},\ }\href {\doibase 10.1103/PhysRevB.69.125313} {\bibfield  {journal}
  {\bibinfo  {journal} {Phys. Rev. B}\ }\textbf {\bibinfo {volume} {69}},\
  \bibinfo {pages} {125313} (\bibinfo {year} {2004})}\BibitemShut {NoStop}%
\bibitem [{\citenamefont {Klatt}\ \emph {et~al.}(2015)\citenamefont {Klatt},
  \citenamefont {M\"uhlbacher},\ and\ \citenamefont
  {Komnik}}]{PhysRevB.91.155306}%
  \BibitemOpen
  \bibfield  {author} {\bibinfo {author} {\bibfnamefont {J.}~\bibnamefont
  {Klatt}}, \bibinfo {author} {\bibfnamefont {L.}~\bibnamefont {M\"uhlbacher}},
  \ and\ \bibinfo {author} {\bibfnamefont {A.}~\bibnamefont {Komnik}},\ }\href
  {\doibase 10.1103/PhysRevB.91.155306} {\bibfield  {journal} {\bibinfo
  {journal} {Phys. Rev. B}\ }\textbf {\bibinfo {volume} {91}},\ \bibinfo
  {pages} {155306} (\bibinfo {year} {2015})}\BibitemShut {NoStop}%
\bibitem [{\citenamefont {Rugar}\ and\ \citenamefont
  {Gr\"utter}(1991)}]{PhysRevLett.67.699}%
  \BibitemOpen
  \bibfield  {author} {\bibinfo {author} {\bibfnamefont {D.}~\bibnamefont
  {Rugar}}\ and\ \bibinfo {author} {\bibfnamefont {P.}~\bibnamefont
  {Gr\"utter}},\ }\href {\doibase 10.1103/PhysRevLett.67.699} {\bibfield
  {journal} {\bibinfo  {journal} {Phys. Rev. Lett.}\ }\textbf {\bibinfo
  {volume} {67}},\ \bibinfo {pages} {699} (\bibinfo {year} {1991})}\BibitemShut
  {NoStop}%
\bibitem [{\citenamefont {Schneider}\ \emph {et~al.}(2012)\citenamefont
  {Schneider}, \citenamefont {Etaki}, \citenamefont {Van Der~Zant},\ and\
  \citenamefont {Steele}}]{schneider2012coupling}%
  \BibitemOpen
  \bibfield  {author} {\bibinfo {author} {\bibfnamefont {B.}~\bibnamefont
  {Schneider}}, \bibinfo {author} {\bibfnamefont {S.}~\bibnamefont {Etaki}},
  \bibinfo {author} {\bibfnamefont {H.}~\bibnamefont {Van Der~Zant}}, \ and\
  \bibinfo {author} {\bibfnamefont {G.}~\bibnamefont {Steele}},\ }\href@noop {}
  {\bibfield  {journal} {\bibinfo  {journal} {Scientific reports}\ }\textbf
  {\bibinfo {volume} {2}} (\bibinfo {year} {2012})}\BibitemShut {NoStop}%
\bibitem [{\citenamefont {Zazunov}\ \emph {et~al.}(2006)\citenamefont
  {Zazunov}, \citenamefont {Feinberg},\ and\ \citenamefont
  {Martin}}]{PhysRevLett.97.196801}%
  \BibitemOpen
  \bibfield  {author} {\bibinfo {author} {\bibfnamefont {A.}~\bibnamefont
  {Zazunov}}, \bibinfo {author} {\bibfnamefont {D.}~\bibnamefont {Feinberg}}, \
  and\ \bibinfo {author} {\bibfnamefont {T.}~\bibnamefont {Martin}},\ }\href
  {\doibase 10.1103/PhysRevLett.97.196801} {\bibfield  {journal} {\bibinfo
  {journal} {Phys. Rev. Lett.}\ }\textbf {\bibinfo {volume} {97}},\ \bibinfo
  {pages} {196801} (\bibinfo {year} {2006})}\BibitemShut {NoStop}%
\bibitem [{\citenamefont {Padurariu}\ \emph {et~al.}(2012)\citenamefont
  {Padurariu}, \citenamefont {Keijzers},\ and\ \citenamefont
  {Nazarov}}]{PhysRevB.86.155448}%
  \BibitemOpen
  \bibfield  {author} {\bibinfo {author} {\bibfnamefont {C.}~\bibnamefont
  {Padurariu}}, \bibinfo {author} {\bibfnamefont {C.~J.~H.}\ \bibnamefont
  {Keijzers}}, \ and\ \bibinfo {author} {\bibfnamefont {Y.~V.}\ \bibnamefont
  {Nazarov}},\ }\href {\doibase 10.1103/PhysRevB.86.155448} {\bibfield
  {journal} {\bibinfo  {journal} {Phys. Rev. B}\ }\textbf {\bibinfo {volume}
  {86}},\ \bibinfo {pages} {155448} (\bibinfo {year} {2012})}\BibitemShut
  {NoStop}%
\end{thebibliography}%

\end{document}